\documentclass{ws-mpla}
\usepackage[super]{cite}
\usepackage{graphicx}
\usepackage{bm}
\usepackage{url}
\usepackage[hidelinks]{hyperref}

\begin{document}

\markboth{E.T. Kipreos \& R.S. Balachandran}
{Rotating frames have a null simultaneity parameter}

\catchline{}{}{}{}{}

\title{Optical data implies a null simultaneity test theory parameter\\in rotating frames\\}

\author{Edward T. Kipreos$^*$ and Riju S. Balachandran$^\dag$}

\address{University of Georgia, 120 Cedar Street, Athens, GA 30602, USA\\
\email{$^*$ekipreos@uga.edu}
\email{$^\dag$balachandranriju@gmail.com}}

\maketitle

\pub{Received 22 December 2020}{Revised 26 April 2021}
\vspace*{-13.5 pt}
\pub{Accepted 6 May 2021}{Published 28 May 2021}

\begin{abstract}
The simultaneity framework describes the relativistic interaction of time with space. The two major proposed simultaneity frameworks are differential simultaneity, in which time is offset with distance in ``moving'' or rotating frames for each ``stationary'' observer, and absolute simultaneity, in which time is not offset with distance. We use the Mansouri and Sexl test theory to analyze the simultaneity framework in rotating frames in the absence of spacetime curvature.  The Mansouri and Sexl test theory has four parameters. Three parameters describe relativistic effects. The fourth parameter, $\epsilon (v)$, was described as a convention on clock synchronization.  We show that $\epsilon (v)$ is not a convention, but is instead a descriptor of the simultaneity framework whose value can be determined from the extent of anisotropy in the unidirectional one-way speed of light.  In rotating frames, one-way light speed anisotropy is described by the Sagnac effect equation.  We show that four published Sagnac equations form a relativistic series based on relativistic kinematics and simultaneity framework.  Only the conventional Sagnac effect equation, and its associated isotropic two-way speed of light, is found to match high-resolution optical data.  Using the conventional Sagnac effect equation, we show that $\epsilon (v)$ has a null value in rotating frames, which implies absolute simultaneity.  Introducing the empirical Mansouri and Sexl parameter values into the test theory equations generates the rotational form of the absolute Lorentz transformation, implying that this transformation accurately describes rotational relativistic effects.

\end{abstract}

\ccode{PACS Nos.: 03.30.+p, 11.30.Cp, 45.40.Bb}

\let\thefootnote\relax\footnote{*Corresponding author.}
\let\thefootnote\relax\footnote{This is an open access article. It is distributed under the terms of the Creative Commons Attribution-NonCommercial-NoDerivs 4.0 (CC BY-NC-ND) License, which permits use and distribution in any medium, provided the original work is properly cited, the use is non-commercial and no modifications or adaptations are made.}

\section{Introduction}\label{section:1}
The Lorentz transformation (LT) describes the relativistic effects of time dilation and length contraction in response to motion \cite{1}.  The LT also describes that time is offset with distance from an observer \cite{1}.  This generates differential simultaneity, also known as the relativity of simultaneity \cite{2, 3}.  With differential simultaneity, events that are simultaneous for one observer will not be simultaneous for another observer in relative motion.  The opposite of differential simultaneity is absolute simultaneity, wherein time is not offset with distance, and events are simultaneous for all observers.  

Relativistic test theories provide a platform to describe the attributes of relativistic transformations \cite{4}. Robust test theories describe both the extent of relativistic effects and the simultaneity framework.  The Mansouri and Sexl (MS) test theory \cite{5, 6, 7} is a robust kinematic test theory that can assess the simultaneity framework.  We recently showed that the MS test theory parameter $\epsilon (v)$ describes the offsetting of time with distance to provide a description of the simultaneity framework \cite{8}.  However, the MS test theory has historically not been used for the analysis of the simultaneity framework because the $\epsilon (v)$ parameter was initially described as a convention on clock synchronization \cite{5}.  The designation as a convention precluded the use of $\epsilon (v)$ in assessing the simultaneity framework.  In Sec. \ref{section:3}, we demonstrate that $\epsilon (v)$ is not a convention, but rather is a descriptor of simultaneity whose value must be empirically determined.  

In contrast to the MS test theory, certain other test theories are not able to assess the simultaneity framework.  For example, the Robertson kinematic test theory implies Einstein (standard) synchronization \cite{4, 9}, which constrains the simultaneity framework to differential simultaneity (see Sec. \ref{section:1.3} and \ref{H}).  The dynamical standard model extension (SME) test theory \cite{10} is also unable to assess the global simultaneity framework because differential simultaneity is ingrained in the underlying structure of the SME.  The SME is not designed to assess \textit{observer} Lorentz invariance (oLI), which addresses how different observers view relativistic effects \cite{11}.  Maintaining oLI implies that all observers make observations from within the LT framework, which includes differential simultaneity.  The SME cannot countenance the complete absence of oLI, as it is built on a framework of oLI \cite{11}.  As an example, boosts and rotations in the SME framework are oLI, and thus imply differential simultaneity \cite{10}.  Further, the observer-invariant nature of the SME\cite{11} precludes a parameter that describes the global simultaneity framework, such as the MS $\epsilon (v)$ parameter. 

\subsection{The LT and ALT transformations and simultaneity}\label{section:1.1}
The four LT equations describe the relativistic effects of time dilation and length contraction in response to motion from the perspective of a ``stationary'' observer: 
\begin{equation}
dt' = \frac{{dt - \frac{{vdx}}{{{c^2}}}}}{{\sqrt {1 - \frac{{{v^2}}}{{{c^2}}}} }},
\label{eq:24}
\end{equation}
\begin{equation}
dx' = \frac{{dx - vdt}}{{\sqrt {1 - \frac{{{v^2}}}{{{c^2}}}} }},
\label{eq:25}
\end{equation}
\begin{equation}
dy' = dy,
\label{eq:26}
\end{equation}
\begin{equation}
dz' = dz,
\label{eq:27}
\end{equation}
where $t$ is the time coordinate, $v$ is velocity, and $x$, $y$, and $z$ are spatial coordinates \cite{1}.  Primed coordinates represent the ``moving'' frame and unprimed coordinates represent the ``stationary'' frame.  The inclusion of a distance term in the LT time coordinate equation (\ref{eq:24}) results in the offsetting of time with distance to generate differential simultaneity \cite{2, 3, 12}.  

Another transformation, ALT, exhibits the same extent of relativistic effects from the stationary perspective as the LT, but lacks a distance term in the time coordinate equation, and so describes absolute simultaneity rather than differential simultaneity \cite{13, 14, 15, 16, 17}.  Other than the time coordinate equation, ALT shares the other equations from the stationary perspective with the LT:
\begin{equation}
dt' = dt\sqrt {1 - \frac{{{v^2}}}{{{c^2}}}},
\label{eq:28}
\end{equation}
\begin{equation}
dx' = \frac{{dx - vdt}}{{\sqrt {1 - \frac{{{v^2}}}{{{c^2}}}} }},
\label{eq:29}
\end{equation}
\begin{equation}
dy' = dy,
\label{eq:30}
\end{equation}
\begin{equation}
dz' = dz.
\label{eq:31}
\end{equation}

ALT has been independently discovered at least six times by: Eagle \cite{18}; Tangherlini \cite{13};  Mansouri and Sexl \cite{5, 6, 7}; Marinov \cite{19}; Vargas \cite{20, 21}; and Selleri \cite{22}.  ALT has been variously called: the absolute Lorentz transformation (ALT) \cite{13}; ether theory \cite{5}; absolute space-time theory \cite{19}; Marinov transformation \cite{21}; inertial transformations \cite{22}; Selleri transformation \cite{23}; Tangherlini transformation (TT) \cite{17}; externally synchronized transformation (EST) \cite{24}; and Lorentz transformation absolute (LTA) \cite{25}.  

Studies of all possible Lorentz-like transformations indicate that only the LT and ALT are capable of matching the classical tests of relativity \cite{21, 26, 27}.  This suggests that the LT and ALT are the only potentially-viable relativistic kinematic transformations.  ALT shares multiple relativistic kinematics with the LT from the stationary perspective, including: time dilation \cite{5, 13, 22}; length contraction \cite{5, 13, 22}; relativistic energy/mass \cite{28}; relativistic Doppler shift \cite{24, 29}; relativistic stellar aberration angle \cite{24, 29}; the one-way speed of light being independent of the velocity of its source \cite{13}; and the isotropic two-way speed of light \cite{13}. These shared kinematics allow the two transformations to generate velocity-invariant null results in experiments with two-way light paths, including Michelson-Morley, Kennedy-Thorndike, and optical resonator experiments \cite{8, 21, 26, 27}, and one-way light paths that are analyzed based on changes in wavelength or frequency \cite{8}.

Both the LT and ALT are compatible with general relativity (GR) in the absence of curvature \cite{13}. The LT is only a unique solution for GR with zero curvature under the stipulation that uniformly translating frames are equivalent in all respects; however, absent this stipulation, ALT is also a solution for GR \cite{13}.  The compatibility of both transformations with GR can be illustrated through their line elements.  The LT ``moving''-frame line element is:
\begin{equation}
ds{'^2} = {c^2}dt{'^2} - dx{'^2} - dy{'^2} - dz{'^2}.
\label{eq:60}
\end{equation}
The ALT moving-frame line element \cite{13, 23} is:
\begin{equation}
ds{'^2} = {c^2}dt{'^2} - 2vdx'dt' - \left( {1 - \frac{{{v^2}}}{{{c^2}}}} \right)dx{'^2} - dy{'^2} - dz{'^2}.
\label{eq:61}
\end{equation}
Substituting the LT equations (\ref{eq:24})--(\ref{eq:27}) into the LT ``moving''-frame line element (\ref{eq:60}) or substituting the ALT equations (\ref{eq:28})--(\ref{eq:31}) into the ALT moving-frame line element (\ref{eq:61}) both give the Minkowski line element, which describes GR in the absence of curvature \cite{12}:
\begin{equation}
d{s^2} = {c^2}d{t^2} - d{x^2} - d{y^2} - d{z^2}.
\label{eq:62}
\end{equation}
Thus, both transformations share the GR ``stationary''-frame line element.

The LT and ALT can be distinguished based on their distinct simultaneity frameworks.  The LT exhibits differential simultaneity, which generates reciprocal relativistic effects that are equivalent between two observers, and isotropic one-way speeds of light (Fig. 1(a)).  In contrast, ALT exhibits absolute simultaneity, which generates directional relativistic effects, so that different observers agree on the directionality of relativistic effects, and anisotropic one-way light speeds for moving observers (Fig. 1(b)).  

\begin{figure}[t]
\centering\includegraphics[width=5.0in]{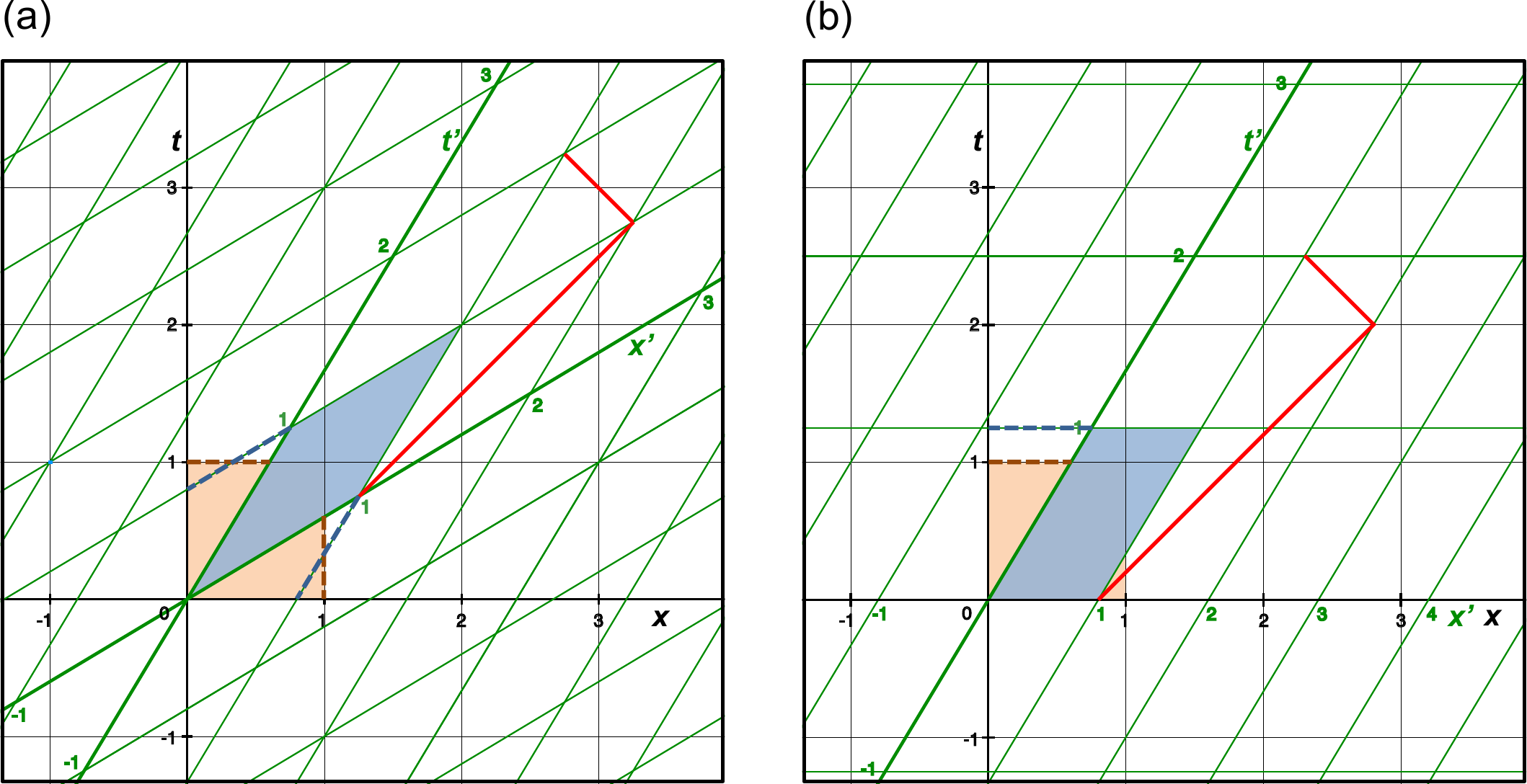}
\caption{Spacetime diagrams with $v = 0.6c$ for the LT (a) and ALT (b). Time is in seconds and distance is in light-seconds. ``Moving'' coordinate lines are green, and ``stationary'' coordinate lines are black. The red line represents a light signal sent from $x' = 1$ to $x' = 2$ where it is reflected back to $x' = 1$. For both transformations, the area of a unit of spacetime is maintained between reference frames -- compare the light-brown square (``stationary'') and the blue rectangle (``moving'') in each diagram. (a) With the LT, time dilation and length contraction are reciprocal. The dashed lines illustrate that times and distances equal to 1.0 in the ``stationary'' frame (dark brown lines) and ``moving'' frame (blue lines) correspond to 0.8 in the other reference frame. (b) With ALT, time dilation and length contraction are directional, as shown with the dark brown and blue dashed lines, and overlaid $x$ and $x'$ axes.}
\label{figure1}
\end{figure}

In the ALT framework, velocities to determine relativistic effects are determined relative to a preferred reference frame (PRF).  Recently, it was recognized that the only type of PRF that allows ALT to be compatible with experimental data are PRFs that are locally associated with gravitational centers \cite{8, 30}.  In the vicinity of the Earth, the corresponding PRF is the non-rotating Earth-centered inertial (ECI) reference frame. The use of the ECI for calculating velocities in the ALT framework is compatible with the widespread use of the ECI as the ``stationary'' reference frame for analyzing relativistic effects in the Earth's local environs.  Because of their identical predictions from the ``stationary'' perspective, the LT and ALT have the same predictions for relativistic experiments analyzed from the ECI perspective \cite{8}.

\subsection{Kinematic analysis of simultaneity}\label{section:1.2}
Our analysis utilizes kinematic data that is insensitive to dynamics: interferometry, the speed of light, doppler shift, and time dilation \cite{31}.  Because the kinematic data does not involve dynamics, it can be fully analyzed by kinematic test theories and transformations.  Our analysis will predominantly use the kinematic MS test theory.

The MS test theory has four parameters \cite{5, 6, 7}: 
\begin{equation}
t' = at + \epsilon x',
\label{eq:20}
\end{equation}
\begin{equation}
x' = b(x - vt),
\label{eq:21}
\end{equation}
\begin{equation}
y' = dy,
\label{eq:22}
\end{equation}
\begin{equation}
z' = dz.
\label{eq:23}
\end{equation}
The extent of relativistic effects are described by $a(v)$ (time dilation), $b(v)$ (length contraction in the direction of motion), and $d(v)$ (length contraction perpendicular to the direction of motion). $\epsilon (v)$ was described as being determined by the convention on clock synchronization \cite{5}.  The LT and ALT share three experimentally-confirmed MS parameter values: $a(v) = (1-v\textsuperscript{2}/c\textsuperscript{2})\textsuperscript{0.5}$; $b(v) = 1/(1-v\textsuperscript{2}/c\textsuperscript{2})\textsuperscript{0.5}$; and $d(v) = 1$ \cite{5, 6, 7}.  For the LT, $\epsilon (v) = -v/c\textsuperscript{2}$, reflecting Einstein synchronization; and for ALT, $\epsilon (v) = 0$, reflecting instantaneous (absolute) synchronization \cite{5}.  We have previously shown that $\epsilon (v)$ describes the simultaneity framework:  $\epsilon (v) = -v/c\textsuperscript{2}$ implies differential simultaneity; and $\epsilon (v) = 0$ implies absolute simultaneity \cite{8}.

Our study will utilize the MS test theory to assess rotational relativistic data.  The original description of the MS test theory described its application to linear inertial reference frames (IRFs) \cite{5, 6, 7}.  However, there was no stipulation that the use of the MS test theory was restricted to IRFs.  The vast majority of relativistic experiments on the Earth or in the local Earth environs constitute rotational data because they either were carried out on the Earth's surface, which is a rotating frame, or in satellites in rotation around the Earth.  The MS test theory has been used extensively to analyze this data, and thus has a sizeable track record in the analysis of rotating-frame data.  

The MS test theory structure allows the identification of the empirical values of its parameters.  Once those empirical values are established, substitution of those values back into the MS test theory equations generates the kinematic relativistic transformation that accurately describes the relativistic data \cite{5}.  

Kinematic rotational transformations (rTs) describe relativistic effects using polar coordinates and peripheral velocity.  The structure of rTs can be identical to that of the corresponding linear transformations except for the substitution of polar coordinates and peripheral velocity.  Examples of this include the rotational form of the LT (the Franklin rT) \cite{32}, and the rotational form of ALT (the ALT rT) \cite{33}.  Kinematic rTs, and the relations that arise from them, can accurately describe relativistic effects using polar coordinates and peripheral velocity.  The rotational time dilation relation, which has the same structure as the linear relation, accurately describes time dilation in response to rotational motion in multiple experimental contexts \cite{34, 35, 36, 37}.  As described below, the predictions of an rT can match experimental data to a resolution of 10\textsuperscript{-18}.  Thus, the rTs and rotational relativistic relations that are expressed with polar coordinates and peripheral velocity accurately describe rotational relativistic data.  The historical usage of the MS test theory demonstrates that it can assess rotational relativistic data; and as we show below, the rotational form of the MS test theory can generate the rT that accurately describes rotational relativistic effects.  Thus, the MS test theory can be used to assess rotational frames.

\subsection{Clock synchronization} \label{section:1.3}
Clock synchronization has historically been considered an important aspect of experimentation when analyzing the one-way speed of light \cite{5, 38}.  Einstein synchronization utilizes forward and backward light signals to synchronize two spatially-separated clocks under the assumption that the speed of light is constant in both directions \cite{1}.  Instantaneous synchronization implies that spatially-separated synchronized clocks are set to the same time in an absolute manner \cite{13}.  

With Einstein synchronization, a ``stationary'' observer will notice that clocks that are spatially separated on a ``moving'' platform are not synchronized in time \cite{2, 12, 39}.  Rather, the clock at the front of the platform has an earlier time than the clock at the back \cite{2, 12, 39}.  From the ``stationary'' perspective, this offsetting of time allows a ``moving''-frame observer to observe that light signals sent from the center of the platform will reach the clocks at the front and back at the same time, thereby allowing the ``moving''-frame observer to calculate isotropic light speeds.  The longer time required for the light to reach the front clock is offset by the earlier timing of the front clock; and the shorter time for light to reach the back clock is offset by the later timing of the back clock \cite{39}.  This offsetting of time manifests as the positive slope of the $x'$-axis (Fig. 1(a)), and reflects differential simultaneity.  Imposing Einstein synchronization on all time coordinates (in combination with time dilation and length contraction) generates LT spacetime \cite{5}.

Conversely, instantaneous synchronization implies a null slope of the $x'$-axis \cite{5} (Fig. 1(b)) in which spatially-separated moving-frame clocks are synchronous for all observers.  Imposing instantaneous synchronization (in combination with time dilation and length contraction) generates ALT spacetime \cite{5}.

Because clock synchronization is associated with a specific simultaneity framework, when we discuss the LT and differential simultaneity then this implies Einstein synchronization, and when we discuss ALT and absolute simultaneity then this implies instantaneous synchronization.  

A number of authors have proposed that Einstein synchronization is not possible in rotating frames, and that instead instantaneous synchronization, corresponding to absolute simultaneity, must be used. Selleri based the conclusion that absolute simultaneity is present in rotating frames on the anisotropy of one-way light speeds in the rotating frame \cite{40, 41}. He noted that symmetry considerations ensure that anisotropic one-way speeds of light are present on all sections of a rotating disk. Selleri proposed a well-known paradox in which the radius of a rotating disk is extended to infinity while maintaining constant peripheral velocity \cite{40, 41}. As the radius is extended toward infinity, the motion of a small section of the disk becomes indistinguishable from that of a linear IRF. In this scenario, either absolute simultaneity with anisotropic one-way speeds of light would be maintained on the small section of the disk or a sharp discontinuity would be required to manifest differential simultaneity and an isotropic one-way speed of light \cite{41}.  

Gift supports the presence of absolute simultaneity in rotating frames by illustrating that the global positioning system is not compatible with differential simultaneity and Einstein synchronization \cite{42}. Lee showed that the application of differential simultaneity and Einstein synchronization to non-rotating cylindrical spacetime generates internal inconsistencies \cite{43}. Spavieri et al. reported that if differential simultaneity with Einstein synchronization is adopted in describing the linear and circular Sagnac effects then light would cover an open path that does not match the closed contour in the measured time interval, leading to a space discontinuity \cite{25, 44}. In these studies, ALT is shown to provide an accurate description without inconsistencies \cite{25, 40, 41, 42, 43, 44}.

Despite the inconsistencies associated with differential simultaneity in rotating frames, in the interest of being comprehensive, we will nevertheless compare the predicted effects of differential simultaneity versus absolute simultaneity in rotating frames.

To ensure that our conclusions do not reflect choices of clock synchronization, we have purposefully limited our analyses to experiments that do not involve clock synchronization, i.e. optical resonators, the classic interferometric Sagnac effect, and ring laser gyroscopes.  These experiments cannot be altered by clock synchronization because they do not utilize clocks.  This eliminates the possibility that the results reflect the choice of clock synchronization.

\section{Overview} \label{section:2}
This study provides multiple new insights.  In Sec. \ref{section:3}, we demonstrate that $\epsilon (v)$ is not a convention.  We will use the one-way light speed anisotropy of the Sagnac effect to determine the value of $\epsilon (v)$ in rotating frames.  In Sec. \ref{section:6}, we demonstrate that the Sagnac effect equation is unchanged from the stationary and rotating perspectives, and that published Sagnac equations form a relativistic series.  In Sec. \ref{section:7}, we show  that the conventional Sagnac effect equation is uniquely compatible with high-resolution optical data.  In Sec. \ref{section:8}, we show that $\epsilon (v)$ can be determined from the anisotropy in the unidirectional one-way speed of light.  Using the anisotropy described by the conventional Sagnac equation, we show that $\epsilon (v)$ has a null value in rotating frames.  In Sec. \ref{section:9}, we demonstrate that the empirical values of the MS parameters generate the ALT rT, implying its compatibility with experimental data.

Throughout, the stationary frame is centered on the non-rotating center of rotation.  The movement of light around a rotating disk is constrained to a circular path along the rim of the disk; this is often considered to arise from the use of multiple mirrors to guide the light or optical fibers.  We use the accepted convention of ``intelligent'' observers, who can take into account the transit times of light \cite{45}.

\section{\boldsymbol{$\epsilon (v)$} is Not a Convention} \label{section:3}
The MS parameters $a(v)$, $b(v)$, and $d(v)$ are descriptors whose values must be experimentally determined \cite{5}.  In contrast, $\epsilon(v)$ was described as a convention that is assigned based on the method used for clock synchronization \cite{5}.  However, our previous analysis suggests that $\epsilon (v)$ is a descriptor, whose value is determined by the simultaneity framework (see Sec. \ref{C}) \cite{8}.  Here, we use the time dilation descriptor $a(v)$ to illustrate four concepts that are relevant for determining whether a parameter is a convention or a descriptor. 

	First, the ability to alter already-collected data to match arbitrarily-assigned values of a parameter does not imply that the parameter is a convention. For example, data can be mathematically converted to reflect a different value of $a(v)$, yet this does not alter that time dilation occurred at the normal relativistic rate.

	Second, the ability to alter an experiment in a finite manner to match an arbitrary value for a parameter does not imply that the parameter is a convention.  For example, clock timing can be altered in a limited manner to match different values of $a(v)$.  Such manipulations are currently instituted for the global positioning system (GPS). GPS satellites are affected by time dilation (due to movement) and time contraction (due to lower gravity) \cite{37}.  These two effects are separable and additive, and, overall, the clocks on the satellites run faster than clocks on the Earth's surface \cite{37, 46}. To correct for this, GPS satellite clock timings are altered so that they match the timing on the Earth's surface \cite{37, 46}. However, the resetting of GPS clocks does not alter that the satellites experience time dilation as a consequence of their velocity relative to the ECI.

	Third, it is not possible to alter experiments in a comprehensive manner to change the value of a descriptor.  For example, while timing can be manipulated in a finite number of GPS satellites, such a change cannot be instituted for all existing clocks.

	Fourth, the value of a descriptor (if it were possible to change it) would affect the outcome of experiments for which the descriptor is important.  For example, the observed null result in a Kennedy-Thorndike experiment depends on time dilation \cite{47}, and so requires the relativistic value of $a(v)$.  It is not possible to assign a different value of $a(v)$ that will generate a non-null result in a Kennedy-Thorndike experiment.  In contrast, a null result in a Michelson-Morley experiment does not depend on time dilation \cite{47}, and so the value of $a(v)$ is irrelevant to the outcome.

	In the following subsections, we will show that $\epsilon(v)$ is not a convention on clock synchronization because: (1) clock resynchronization to switch between $\epsilon(v)$ values is not possible due to internal inconsistencies that arise with two or more observers; and (2) the value of $\epsilon(v)$ is critical in determining the outcome of experiments that rely on it, but the values cannot be arbitrarily assigned to alter the outcomes of those experiments.

\subsection{Assessing the interconvertibility of simultaneity frameworks} \label{section:3.1}
Within the MS framework, the LT and ALT only differ for their $\epsilon(v)$ parameter value.  If $\epsilon(v)$ were a convention then it should be possible to assign a value of $\epsilon(v)$ that would interconvert ALT and the LT.  In that vein, it has been proposed that the LT and ALT are alternative forms that can be interconverted by changing clock synchronization gauges \cite{48}.  

Multiple studies have identified invalidating inconsistencies when the differential simultaneity of the LT is applied to rotating or spatially-closed systems \cite{23, 25, 33, 42, 43, 44}.  These studies suggest that the LT and ALT describe different physical realities that are not interchangeable because only ALT is compatible with such spatially-closed systems, while the LT is not.  Nevertheless, in this section, we will determine whether resynchronizations to interconvert the two transformations are physically possible based on the ability to satisfy the expectations of two observers in relative motion. 

The physical mechanism to interconvert the LT and ALT would be to resynchronize ``moving'' clocks.  As viewed by a ``stationary'' observer, the time offset with distance associated with differential simultaneity is referred to as the relativistic time offset ($RTO$) \cite{8}.  For a linear IRF, the $RTO$ (for forward and reverse directions) \cite{2} is:
\begin{equation}
RTO = \mp \frac{{vl'}}{{{c^2}}},
\label{eq:75}
\end{equation}
where $l'$ is the length between two ``moving''-frame points.  For an observer at the origin of a spacetime diagram, $l'$ is equivalent to $x'$ coordinates, generating $RTO = \mp vx'/c\textsuperscript{2}$. Along with the definition of the $RTO$ given in Eq. (\ref{eq:75}), the $RTO$ can also be defined as $RTO = \mp vl/c\textsuperscript{2}$, if $l$ is defined as the ``at rest'', ``stationary'' length of the ``moving''-frame platform \cite{8}. 

Here, we show that ALT can be mathematically converted to the LT by adding the $RTO$ to all moving-frame ALT clock timings.  Conversely, the LT can be mathematically converted to ALT by removing the $RTO$ from every ``moving''-frame LT clock timings.  To convert ALT to the LT, one adds the $RTO$ to the ALT $t'$ equation (\ref{eq:28}), and with the substitution of the value of the shared ALT/LT $x'$ equations (\ref{eq:25}) and (\ref{eq:29}), this gives the LT $t'$ equation (\ref{eq:24}):
\begin{equation}
{t'_{{\rm{ALT to LT}}}} = t\sqrt {1 - \frac{{{v^2}}}{{{c^2}}}}  - \frac{{vx'}}{{{c^2}}} = \frac{{t - \frac{{vx}}{{{c^2}}}}}{{\sqrt {1 - \frac{{{v^2}}}{{{c^2}}}} }}.
\label{eq:130}
\end{equation}
Similarly, removing the $RTO$ from the LT $t'$ equation (\ref{eq:24}) gives the ALT $t'$ equation (\ref{eq:28}):
\begin{equation}
{t'_{{\rm{LT to ALT}}}} = \frac{{t - \frac{{vx}}{{{c^2}}}}}{{\sqrt {1 - \frac{{{v^2}}}{{{c^2}}}} }} + \frac{{vx'}}{{{c^2}}} = t\sqrt {1 - \frac{{{v^2}}}{{{c^2}}}} .
\label{eq:131}
\end{equation}

The interconversion of the $t'$ equations has the effect of interconverting the transformations, as the transformations only differ in their $t'$ equations.  This demonstrates that the LT and ALT share the same basic structure, and differ only in the presence or absence of differential simultaneity.  

As illustrated above, it is possible to \textit{mathematically} interconvert the two transformations by clock resynchronizations.  However, in the LT framework, the offsetting of time with distance is set for each observer based on the velocity of the object that is being observed, with the magnitude of the time offset changing based on the position of the object \cite{2}.  Therefore, the time on a clock in the ``moving'' frame, as viewed by a ``stationary'' observer, is dependent on the velocity and distance of the clock from the observer.  This requirement to resynchronize clocks based on the distance to each observer affects the resynchronizations between the LT and ALT in both directions.  For an ALT reality to be resynchronized to the LT, all observed clocks have to be resynchronized to incorporate the time offset based on their velocity and distance from the observer (to match the expectations of the LT).  For an LT reality to be resynchronized to ALT, all clocks observed have to be resynchronized to remove the time offset based on the clock's velocity and distance from the observer (as the naturally-occurring time offset of the LT has to be removed from each clock that is viewed by the observer).  One can envision the practical impossibility of experimentally altering all clocks that an observer views based on their changing distance from the observer.  This task is compounded by the fact that not all clocks can be readily manipulated, such as the half-lives of radioactive compounds.  

In any relativistic experiment based on motion it is necessary to include, at a minimum, two reference frames in order to allow a comparison of relativistic effects between them.  As we will show below, when observers are present in both reference frames being compared, there is an internal inconsistency that prevents the resynchronization of clocks that satisfy one observer from simultaneously satisfying the expectations of the other observer.  This is because there is a disconnect between the clock resynchronization and the underlying simultaneity of spacetime that affects the observations of the second observer.  Here we will consider ``stationary'' and ``moving'' observers instantaneously coincident at position (0,0).  Before we describe their observations after resynchronizations, we first need to discuss how the two observers' IRFs are offset in time relative to each other in the LT framework.

In a differential simultaneity reality, a ``moving'' platform is shifted in time over its length relative to the ``stationary'' frame (Fig. 2(a)).  At one instant of ``moving'' time, the platform spans different times in the ``stationary'' frame.  At time 0, a ``stationary'' observer at position (0,0) observes the ``moving'' platform as the transparent-yellow outline in Fig. 2(a), which is comprised of different slices of ``moving''-frame time, shown as light-shaded (earlier) to dark-shaded (later) ``moving''-frame platforms. The ``stationary'' observer observes that the platform (yellow outline) has different ``moving''-frame times over its length \cite{39}.  The back end of the platform has time 0, while the front end of the platform has the earlier ``moving''-frame time of $-vl'/c\textsuperscript{2}$ due to the $RTO$ (Fig. 2(a), red lettering) \cite{2, 8}. In contrast, a ``moving''-frame observer at position (0,0) observes synchronous time 0 along the length of the dark-shaded ``moving''-frame platform (Fig. 2(a), blue lettering).  

\begin{figure}[!htbp]
\centering\includegraphics[width=4.8in]{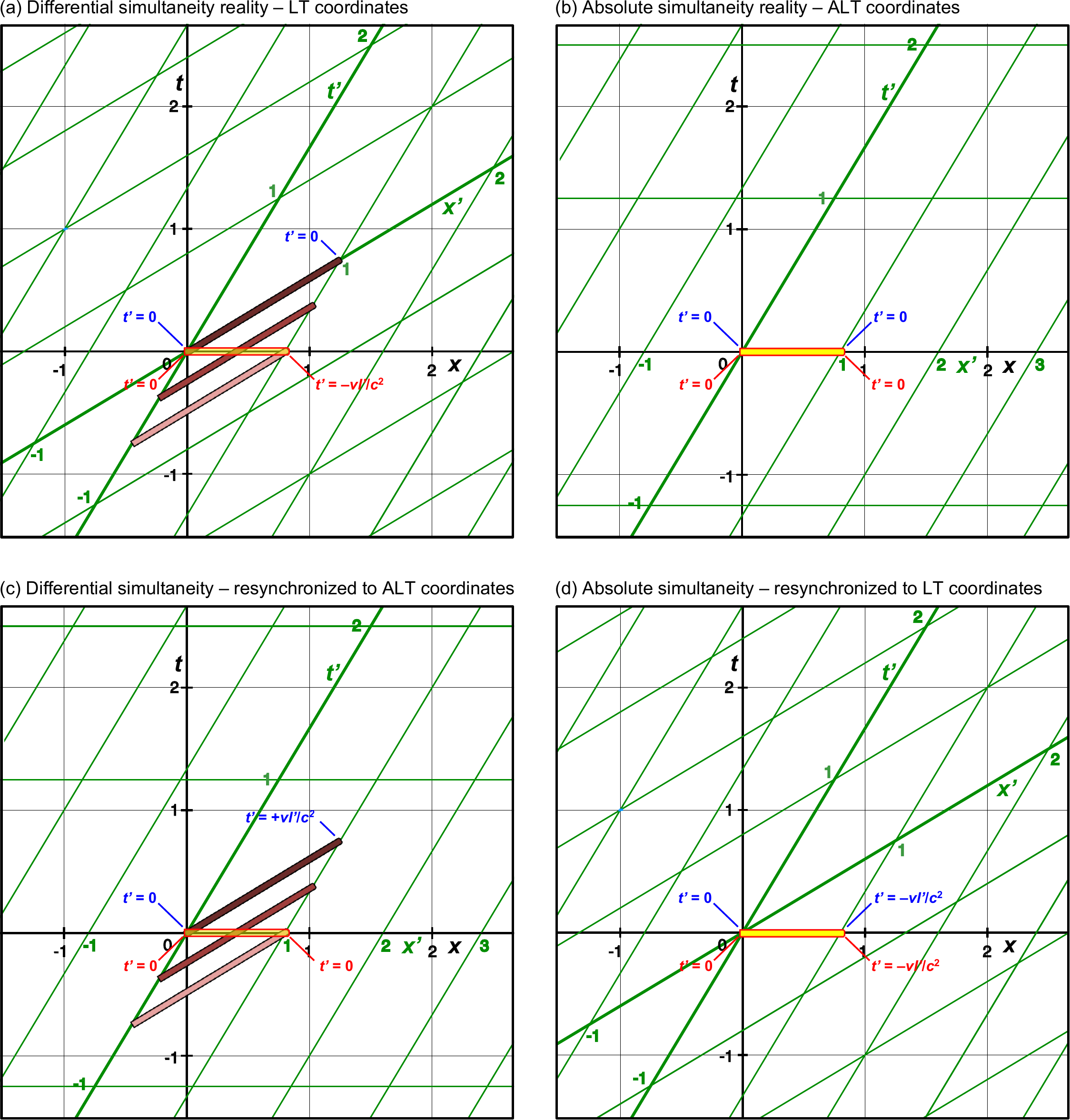}
\caption{Resynchronization of LT and ALT coordinates in differential and absolute simultaneity realities.  Spacetime diagrams with $v = 0.6c$ show a ``moving'' platform of length $l' = 1$ with the back end at coordinate (0,0).  ``Moving''- and ``stationary''-frame (0,0) observers view ``moving''-frame clocks at the back and front ends of the platform, with the clock times reported in blue for the ``moving''-frame observer, and in red for the ``stationary''-frame observer.  (a) A differential simultaneity reality with LT coordinates.  (b) An absolute simultaneity reality with ALT coordinates.  (c) A differential simultaneity reality in which ``moving''-frame time coordinates have been altered by the factor $+vx'/c\textsuperscript{2}$ to generate the ALT coordinate system.  The platform is still offset in time due to the effects of differential simultaneity.  (d) An absolute simultaneity reality in which moving-frame time coordinates have been altered by the factor $-vx'/c\textsuperscript{2}$ to generate the LT coordinate system.  The platform remains non-time-offset. For both resynchronizations (c) and (d), the ``moving''-frame observer no longer views synchronous clock timing on the ``moving'' platform.}  
\label{figure2}
\end{figure}

In an absolute simultaneity reality, platforms are not offset in time with distance.  Therefore, clocks at the front and back of a moving platform have synchronized times for both stationary- and moving-frame observers (Fig. 2(b), red and blue lettering).

We now consider what happens during clock resynchronizations.  In a differential simultaneity reality, resynchronizing ``moving’’-frame time coordinates by $+vx'/c\textsuperscript{2}$ converts the LT coordinate system (Fig. 2(a)) to the ALT coordinate system (Fig. 2(c)).  Correspondingly, in an absolute simultaneity reality, resynchronizing moving-frame time coordinates by $-vx'/c\textsuperscript{2}$ converts the ALT coordinate system (Fig. 2(b)) to the LT coordinate system (Fig. 2(d)).  However, there is no known mechanism that can interconvert the underlying reality from differential simultaneity to absolute simultaneity or vice versa.  Because the underlying simultaneity frameworks cannot be interconverted, objects in a differential simultaneity reality would remain offset in time after resynchronization, while objects in an absolute simultaneity reality would remain not offset in time.  The disconnect between the resynchronized coordinates and the underlying simultaneity generates internal inconsistencies.  

Consider a differential simultaneity reality in which LT time coordinates are resynchronized to ALT time coordinates (Fig. 2(c)).  A ``stationary''-frame observer at position (0,0) would observe that the ``moving''-frame clock at the front of the platform reads 0, as expected for ALT.  However, a ``moving''-frame (0,0) observer would expect the clock at the front of the platform to synchronously read 0, but it would instead read $+vl'/c\textsuperscript{2}$.  Conversely, in an absolute simultaneity reality in which ALT coordinates are resynchronized to LT coordinates, a stationary-frame (0,0) observer would observe the clock at the front of the platform to read $-vl'/c\textsuperscript{2}$, as expected for the LT (Fig. 2(d)).  However, a moving-frame (0,0) observer would expect the clock at the front of the platform to synchronously read 0, but it would instead read $-vl'/c\textsuperscript{2}$.  Thus, interconversions of simultaneity frameworks would not allow even two observers to experience the resynchronized framework without internal inconsistencies.

\subsection{$\epsilon (v)$ values alter the Sagnac effect}\label{section:3.2}
It has been proposed that $\epsilon (v)$ is a convention because it could be eliminated from MS test theory-based descriptions of four different experiments that analyzed wavelength and frequency in one-way light paths \cite{49}.  However, the four experiments are incapable of distinguishing the LT and ALT because of shared relativistic kinematics that generate null results for each experiment \cite{8, 50, 51, 52}.  Because the LT and ALT only differ for the $\epsilon (v)$ parameter, experiments that cannot distinguish the two transformations are not optimal for analyzing the importance of $\epsilon (v)$.  A more informative approach is to analyze experiments that can distinguish the LT and ALT.  

Here we show that the value of $\epsilon (v)$ is critical for determining the generalized Sagnac effect.  The generalized Sagnac effect is based on the observation that fiber-optic light paths with majority linear segments exhibit the Sagnac effect in proportion to their total length, irrespective of the proportion that is linear \cite{53, 54}.  The basic Sagnac effect equation from the rotating-frame perspective is the difference in timing between the co-rotating and counter-rotating light signals emitted and received by the moving emitter/receiver:
\begin{equation}
{\rm{Sagna}}{{\rm{c}}_{{\rm{Rotating}}}} = t{'_{{\rm{co - rot}}}} - t{'_{{\rm{counter - rot}}}} = \frac{{l'}}{{c{'_{{\rm{co - rot}}}}}} - \frac{{l'}}{{c{'_{{\rm{counter - rot}}}}}}.
\label{eq:38}
\end{equation}
The rotating-frame arc length of light propagation ($l'$) in each direction is the same, but the rotating-frame speed of light ($c'$) is different in each direction.

To express Eq. (\ref{eq:38}) in MS test theory parameters, the one-way speed of light in rotating frames is given by $c' = dx'/dt'$, with $dx'$ and $dt'$ referring to the motion of light. $dx'$ is defined by the MS equation (\ref{eq:21}).  $dt'$ is defined by the MS equation (\ref{eq:20}) for which $dx'$ is substituted with the value from (\ref{eq:21}).  This produces the generalized Sagnac equation described with MS parameters:
\begin{equation}
{\rm{Sagnac}} = \frac{{l'}}{{\left( {\frac{{b(dx - vdt)}}{{adt + \epsilon b(dx - vdt)}}} \right)}} - \frac{{l'}}{{\left( {\frac{{b(dx + vdt)}}{{adt - \epsilon b(dx + vdt)}}} \right)}}.
\label{eq:63}
\end{equation}
Simplifying Eq. (\ref{eq:63}), and converting $dx/dt$ to $c$ gives:
\begin{equation}
{\rm{Sagnac}} = \frac{{2l'va}}{{b{c^2}\left( {1 - \frac{{{v^2}}}{{{c^2}}}} \right)}} + 2l'\epsilon.
\label{eq:64}
\end{equation}

We will determine the Sagnac equations that correspond to the LT and ALT by incorporating the MS parameter values associated with each transformation into Eq. (\ref{eq:64}).  Substituting the ALT MS-parameter values (absolute simultaneity, $\epsilon (v) = 0$; time dilation, $a(v) = (1 - v\textsuperscript{2}/c\textsuperscript{2})\textsuperscript{0.5}$; and length contraction, $b(v) = 1/(1 - v\textsuperscript{2}/c\textsuperscript{2})\textsuperscript{0.5}$) gives the conventional Sag\textsubscript{2AS} equation (named for 2 relativistic effects and absolute simultaneity):
\begin{equation}
{\rm{Sa}}{{\rm{g}}_{2{\rm{AS}}}} = \frac{{2l'v\sqrt {1 - \frac{{{v^2}}}{{{c^2}}}} \sqrt {1 - \frac{{{v^2}}}{{{c^2}}}} }}{{{c^2}\left( {1 - \frac{{{v^2}}}{{{c^2}}}} \right)}} = \frac{{2l'v}}{{{c^2}}}.
\label{eq:68}
\end{equation}

Substituting the LT MS-parameter values (differential simultaneity, $\epsilon (v) = -v/c\textsuperscript{2}$; time dilation; and length contraction) gives Sag\textsubscript{2DS} (named for differential simultaneity).  The null value reflects the isotropic one-way speed of light:
\begin{equation}
{\rm{Sa}}{{\rm{g}}_{{\rm{2DS}}}} = \frac{{2l'v\sqrt {1 - \frac{{{v^2}}}{{{c^2}}}} \sqrt {1 - \frac{{{v^2}}}{{{c^2}}}} }}{{{c^2}\left( {1 - \frac{{{v^2}}}{{{c^2}}}} \right)}} - \frac{{2l'v}}{{{c^2}}} = 0.
\label{eq:69}
\end{equation}

These Sagnac equations can also be derived for rotating frames by utilizing the ALT rT and Franklin rT MS parameter values, which are expressed with peripheral velocity ($\omega r$) and polar coordinates, and a light path of one circumference ($2 \pi r$).  The ALT rT has the MS parameter values: $a(v) = (1 - \omega\textsuperscript{2} r\textsuperscript{2}/c\textsuperscript{2})\textsuperscript{0.5}$; $b(v) = 1/(1 - \omega\textsuperscript{2} r\textsuperscript{2} /c\textsuperscript{2})\textsuperscript{0.5}$; and $\epsilon (v) = 0$. The Franklin rT shares the $a(v)$ and $b(v)$ values, but has $\epsilon (v) = -\omega r/c\textsuperscript{2}$. Substituting these values into Eq. (\ref{eq:64}) generates the Sagnac equations for the ALT rT and Franklin rT, respectively:
\begin{equation}
{\rm{Sa}}{{\rm{g}}_{2{\rm{AS}}}} = \frac{{2(2\pi r)\omega r\sqrt {1 - \frac{{{\omega ^2}{r^2}}}{{{c^2}}}} \sqrt {1 - \frac{{{\omega ^2}{r^2}}}{{{c^2}}}} }}{{{c^2}\left( {1 - \frac{{{\omega ^2}{r^2}}}{{{c^2}}}} \right)}} = \frac{{4\pi \omega {r^2}}}{{{c^2}}},
\label{eq:77}
\end{equation}
\begin{equation}
{\rm{Sa}}{{\rm{g}}_{{\rm{2DS}}}} = \frac{{2(2\pi r)\omega r\sqrt {1 - \frac{{{\omega ^2}{r^2}}}{{{c^2}}}} \sqrt {1 - \frac{{{\omega ^2}{r^2}}}{{{c^2}}}} }}{{{c^2}\left( {1 - \frac{{{\omega ^2}{r^2}}}{{{c^2}}}} \right)}} - \frac{{2(2\pi r)\omega r}}{{{c^2}}} = 0.
\label{eq:78}
\end{equation}

The Sagnac equations (\ref{eq:77}) and (\ref{eq:78}) are equivalent to those obtained by deriving the equations directly from the ALT rT and Franklin rT transformation equations, respectively (see our companion study \cite{55}). The ALT rT and Franklin rT Sagnac equations are qualitatively different.  Incorporating $\epsilon (v) = -\omega r/c\textsuperscript{2}$ generates a null Sagnac effect (\ref{eq:78}) that is velocity invariant, while incorporating $\epsilon (v) = 0$ generates the conventional Sagnac effect (\ref{eq:77}) that is a function of peripheral velocity.  These are fundamentally different effects, with only the latter experimentally observed \cite{56}. 

That the $\epsilon (v)$ parameter is critical for determining the Sagnac effect equation provides support for the position that $\epsilon (v)$ is not a convention on clock synchronization.  Both the conventional and generalized interferometric Sagnac effect experiments lack clocks, and thus $\epsilon (v)$ cannot function to synchronize clocks in these experiments. As we have described, $\epsilon (v)$ describes the offsetting of time with distance \cite{8}, and thus the value of $\epsilon (v)$ is required for the Sagnac effect because it describes the spatial distribution of rotating-frame time along the disk. 

A recent paper analyzed the generalized Sagnac effect using the MS test theory \cite{57}.  This paper reported that the LT $\epsilon (v) = -v/c\textsuperscript{2}$ generates the Sag\textsubscript{2AS} equation.  This is in contrast to our analysis that shows that $\epsilon (v) = -v/c\textsuperscript{2}$ generates the null Sag\textsubscript{2DS}, while the ALT $\epsilon (v) = 0$ generates Sag\textsubscript{2AS}.  The paper treated the $a(v)$, $b(v)$, and $d(v)$ parameters as scalars.  $\epsilon (v)$ was designated as a vector despite being described as a convention that was fixed by the synchronization scheme.  As described in \ref{C}, $\epsilon (v)$ describes the offsetting of time with distance.  This implies that it is a scalar whose value is fixed by the simultaneity framework and is independent of the direction of motion.  Nevertheless, the decision to analyze $\epsilon (v)$ as a vector did not generate the discrepancy with our results.  Rather, the discrepancy arose from an error in the choice of the LT $dt'$ equation that was used to generate $c'$.  

The error arose from using the shared LT/ALT time relation value of $dt'$ ($dt'=dt(1-v\textsuperscript{2}/c\textsuperscript{2})\textsuperscript{0.5}$) \cite{1, 5, 13} to calculate $c' = dx'/dt'$, rather than the LT $dt'$ transformation equation (\ref{eq:24}).  Critically, the shared time dilation relation is identical to the ALT $dt'$ transformation equation (\ref{eq:28}), but is distinct from the LT $dt'$ transformation equation (\ref{eq:24}).  Moreover, the LT and ALT share the same $dx'$ transformation equation (\ref{eq:25} and \ref{eq:29}).  Therefore, the analysis was equivalent to calculating the ALT $c'$ value by dividing the ALT $dx'$ equation by the ALT $dt'$ equation.  The ALT $c'$ value was then used to obtain the ALT-associated conventional Sag\textsubscript{2AS} equation \cite{57}. 

\subsection{The inability to arbitrarily assign $\epsilon (v)$ values for certain experiments} \label{section:3.3}
A major criterion of a convention is that it can be chosen arbitrarily.  However, it is not possible to alter the value of $\epsilon (v)$ to alter experiments that do not utilize physical clocks, such as the classic Sagnac effect interferometry experiment.  Even the alteration of $\epsilon (v)$ in experiments with conventional clocks can only be carried out in a limited manner via adjustment of a finite number of clocks from restricted perspectives (e.g. the ``stationary'' frame).  This is because it is not possible to resynchronize clock timing to accommodate ``moving'' observers that are observing clock times in different reference frames (see Sec. \ref{section:3.1}).  

	$\epsilon (v)$ is thus not a convention on clock synchronization because it cannot be experimentally instituted for multiple observers without introducing internal inconsistencies, and its value cannot be chosen to alter experiments that depend on it.  For example, it is not possible to assign $\epsilon (v) = -v/c\textsuperscript{2}$ in a generalized Sagnac interferometry experiment and generate a null Sagnac effect that is invariant to velocity.

\section{Sagnac Equations Form a Relativistic Series} \label{section:6}

We will use the anisotropy in the rotating-frame speed of light to determine the value of $\epsilon (v)$.  In 1925, Michelson and Gale showed that light speed is anisotropic on the surface of the rotating Earth \cite{58, 59}, and this is routinely demonstrated today \cite{60}.  In rotating frames, one-way light speed anisotropy is described by the Sagnac effect equation \cite{61}.  However, before we can use the Sagnac effect to calculate the value of $\epsilon (v)$, we must determine which of several Sagnac equations is validated by experimental data.  Prior to that analysis, we will demonstrate that the published Sagnac equations: 1) are equivalent from the rotating- and stationary-frame perspectives; 2) are associated with specific combinations of relativistic attributes; and 3) have specific one-way and two-way speeds of light.  

This description of how Sagnac equations are linked to relativistic conditions is important because multiple papers in the literature state that a variant Sag\textsubscript{1AS} equation is the valid Sagnac equation, e.g. see (\cite{62, 63, 64, 65, 66, 67}).  Sag\textsubscript{1AS} has been referred to as the relativistic form of the Sagnac equation \cite{62, 65}.  However, we show that Sag\textsubscript{1AS} is only partially relativistic, while Sag\textsubscript{2AS} and Sag\textsubscript{2DS} are fully relativistic (for absolute and differential simultaneity, respectively).  It is also important to clarify that the Sagnac equation has the same form for both stationary and rotating frames.  This ensures that a given combination of relativistic conditions corresponds to a single Sagnac equation irrespective of perspective.

\subsection{Stationary- and rotating-frame Sagnac equations are equivalent} \label{section:6.1}
Here we describe the basic Sagnac effect from the rotating-frame and stationary-frame perspectives using a non-relativistic (Galilean) framework that lacks relativistic effects and exhibits absolute simultaneity. This analysis will show that the Sagnac equations from the two perspectives are identical.

In the classic Sagnac effect experiment, a rotating emitter/receiver sends light signals in co-rotating and counter-rotating directions.  The interference pattern is a function of the different timings of light propagation in the two directions.  At any given instant, the light that strikes the receiver from the co-rotating direction was emitted earlier than the light that strikes the receiver from the counter-rotating direction.  From the stationary perspective, both light signals propagate at the same isotropic speed $c$, but the co-rotating light traverses a longer arc length to the receiver than the counter-rotating light (see Fig. \ref{FigC.1} in \ref{A}).  From the rotating-frame perspective, the arc lengths are the same, but the co-rotating light propagates slower than the counter-rotating light. 

The rotating-frame perspective is that of an observer on a rotating disk.  This is equivalent to an observer on the rotating surface of the Earth that considers east--west distances.  The rotating Earth observer considers the distances on a map to be fixed, with a distance of 10 km to the east being equivalent to a distance of 10 km to the west.  If the rotating-frame observer sends a light signal to the east, and a light signal to the west, the light will take more time to the reach the eastward 10 km mark (in the co-rotating direction) than to reach the westward 10 km mark (in the counter-rotating direction). This has been demonstrated by Michelson and Gale \cite{58, 59}, and by Sagnac effect experiments that are able to measure the Earth's speed of rotation \cite{60}.  It has been experimentally demonstrated that light signals that propagate around the world from an emission/reception point on the Earth's surface take more time to propagate around the world when the light signal is sent in the eastward direction, and less time when the light signal is sent in the westward direction \cite{68}.   From the rotating-frame perspective, the circumference is $2 \pi r$ regardless of whether a light signal is sent in the co-rotating or counter-rotating direction.  Thus, a calculation of rotating-frame distance divided by time indicates that the rotating-frame one-way speed of light is slower than $c$ in the co-rotating direction, and greater than $c$ in the counter-rotating direction. In the non-relativistic Galilean framework,  $c'\textsubscript{co-rot} = c - \omega r$, and $c'\textsubscript{counter-rot} = c + \omega r$ \cite{69}.  

Equation (\ref{eq:38}) shows the basic Sagnac equation from the rotating-frame perspective.  The distance for light propagation in both directions is the circumference of $2 \pi r$.  Substituting this distance and the Galilean $c'$ values into Eq. (\ref{eq:38}) generates the non-relativistic, Galilean Sagnac equation from the rotating-frame perspective \cite{69}:
\begin{equation}
{\rm{Sa}}{{\rm{g}}_{{\rm{0AS}}}} = \frac{{2\pi r}}{{c - \omega r}} - \frac{{2\pi r}}{{c + \omega r}} = \frac{{4\pi \omega {r^2}}}{{{c^2}\left( {1 - \frac{{{\omega ^2}{r^2}}}{{{c^2}}}} \right)}}.
\label{eq:39}
\end{equation}

From the stationary perspective, the speed of light is isotropic $c$ in both co-rotating and counter-rotating directions, but the distances are different.  The speed of light is isotropic $c$ because the stationary perspective is that of the PRF, where the speed of light is $c$ \cite{5}.  In the co-rotating direction, the emitter/receiver rotates away from the light signal, making the arc length of light propagation greater than one circumference (Fig. \ref{FigC.1}). Conversely, in the counter-rotating direction, the emitter/receiver rotates toward the light signal, making the arc length less than one circumference (Fig. \ref{FigC.1}). The basic equation for the Sagnac effect from the stationary perspective is:
\begin{equation}
{\rm{Sagna}}{{\rm{c}}_{{\rm{Stationary}}}} = {t_{{\rm{co - rot}}}} - {t_{{\rm{counter - rot}}}} = \frac{{{l_{{\rm{co - rot}}}}}}{c} - \frac{{{l_{{\rm{counter - rot}}}}}}{c}.
\label{eq:98}
\end{equation}

From the Galilean stationary perspective, the longer distance that light propagates in the co-rotating direction is $2 \pi r$ multiplied by the ratio $c$/$c'\textsubscript{co-rot}$ (see \ref{A} for the derivation).  The shorter distance that light propagates in the counter-rotating direction is $2 \pi r$ multiplied by the ratio $c$/$c'\textsubscript{counter-rot}$ (\ref{A}).  Incorporating these distances in the basic stationary-frame Sagnac equation (\ref{eq:98}) produces the Galilean Sagnac equation from the stationary-frame perspective:
\begin{equation}
{\rm{Sa}}{{\rm{g}}_{{\rm{0AS}}}} = \frac{{2\pi r\left( {\frac{c}{{c - \omega r}}} \right)}}{c} - \frac{{2\pi r\left( {\frac{c}{{c + \omega r}}} \right)}}{c} = \frac{{4\pi \omega {r^2}}}{{{c^2}\left( {1 - \frac{{{\omega ^2}{r^2}}}{{{c^2}}}} \right)}},
\label{eq:50}
\end{equation}

Note that the Galilean Sagnac effect equations from the rotating-frame (\ref{eq:39}) and the stationary-frame (\ref{eq:50}) perspectives are identical.  The timing differences that generate the Sagnac effect can produce discrete physical events, such as interference patterns.  These are frame-independent physical events, and all observers agree on the nature of such events \cite{70}. Because all observers agree on the exact interference pattern that is produced, they also agree on the associated Sagnac effect equation.  The incorporation of relativistic effects and different simultaneity frameworks does not alter the equivalence of the rotating-frame and stationary-frame Sagnac equations, as shown in Sec. \ref{section:6.2}.

\subsection{A relativistic series of Sagnac equations} \label{section:6.2}
Here, the Sagnac equations are derived in a relativistic series based on the incorporation of relativistic effects and simultaneity frameworks.  When introducing relativistic effects into the Sagnac equation, the critical aspect is how the timing of light propagation is altered.  Thus, length contraction reduces the overall length of the light propagation, which reduces the timing.  Time dilation similarly reduces the timing in the rotating frame.  

From the rotating perspective, the Sagnac equations show the rotating-frame one-way speeds of light in the denominators of the timing expressions (see Eq. \ref{eq:38}).  A specific set of co-rotating and counter-rotating one-way speeds of light is sufficient to uniquely specify a Sagnac equation.  Incorporation of a given combination of relativistic effects into the rotating-frame Galilean Sagnac equation (\ref{eq:39}) and rearrangement into the denominator shows the one-way speeds of light for that Sagnac effect.  The rotating-frame one-way speeds of light for a Sagnac effect with a given set of relativistic effects are equivalent to the rotating-frame one-way speeds of light of the rT that exhibits the same combination of relativistic effects (see our companion study \cite{55}).

Sag\textsubscript{1AS} is obtained by incorporating one relativistic effect (time dilation or length contraction) into Eq. (\ref{eq:39}).  Rearranging the relativistic term into the denominator reveals the rotating-frame one-way speeds of light (in the middle expression) and Sagnac equation associated with the Post rT \cite{55, 71}, which exhibits only time dilation:
\begin{eqnarray}
{\rm{Sa}}{{\rm{g}}_{{\rm{1AS}}}} &=& \frac{{2\pi r\sqrt {1 - \frac{{{\omega ^2}{r^2}}}{{{c^2}}}} }}{{c - \omega r}} - \frac{{2\pi r\sqrt {1 - \frac{{{\omega ^2}{r^2}}}{{{c^2}}}} }}{{c + \omega r}} \nonumber \\
 &=& \frac{{2\pi r}}{{\left( {\frac{{c - \omega r}}{{\sqrt {1 - \frac{{{\omega ^2}{r^2}}}{{{c^2}}}} }}} \right)}} - \frac{{2\pi r}}{{\left( {\frac{{c + \omega r}}{{\sqrt {1 - \frac{{{\omega ^2}{r^2}}}{{{c^2}}}} }}} \right)}} 
= \frac{{4\pi \omega {r^2}}}{{{c^2}\sqrt {1 - \frac{{{\omega ^2}{r^2}}}{{{c^2}}}} }}.
\label{eq:40}
\end{eqnarray}

Sag\textsubscript{1AS} can also be obtained by incorporating time dilation, length contraction, and a larger circumference of $2 \pi r/(1 - \omega \textsuperscript{2}r\textsuperscript{2}/c\textsuperscript{2})\textsuperscript{0.5}$ within the rotating frame, which are the conditions of the Langevin metric \cite{72, 73, 74}.  This reveals the one-way speeds of light and Sagnac equation associated with the Langevin metric \cite{75, 76, 77}:
\begin{eqnarray}
{\rm{Sa}}{{\rm{g}}_{{\rm{1AS}}}} &=& \frac{{\frac{{2\pi r}}{{\sqrt {1 - \frac{{{\omega ^2}{r^2}}}{{{c^2}}}} }}\left( {1 - \frac{{{\omega ^2}{r^2}}}{{{c^2}}}} \right)}}{{c - \omega r}} - \frac{{\frac{{2\pi r}}{{\sqrt {1 - \frac{{{\omega ^2}{r^2}}}{{{c^2}}}} }}\left( {1 - \frac{{{\omega ^2}{r^2}}}{{{c^2}}}} \right)}}{{c + \omega r}} 
\nonumber \\
&=& \frac{{2\pi r}}{{\left( {\frac{{c - \omega r}}{{\sqrt {1 - \frac{{{\omega ^2}{r^2}}}{{{c^2}}}} }}} \right)}} - \frac{{2\pi r}}{{\left( {\frac{{c + \omega r}}{{\sqrt {1 - \frac{{{\omega ^2}{r^2}}}{{{c^2}}}} }}} \right)}} = \frac{{4\pi \omega {r^2}}}{{{c^2}\sqrt {1 - \frac{{{\omega ^2}{r^2}}}{{{c^2}}}} }}.
\label{eq:43}
\end{eqnarray}

Sag\textsubscript{2AS} is obtained by incorporating both time dilation and length contraction into Eq. (\ref{eq:39}), which are the conditions of the ALT rT \cite{33}.  This reveals the one-way speeds of light and Sagnac equation associated with the ALT rT:\cite{55, 78}
\begin{eqnarray}
{\rm{Sa}}{{\rm{g}}_{{\rm{2AS}}}} &=& \frac{{2\pi r\left( {1 - \frac{{{\omega ^2}{r^2}}}{{{c^2}}}} \right)}}{{c - \omega r}} - \frac{{2\pi r\left( {1 - \frac{{{\omega ^2}{r^2}}}{{{c^2}}}} \right)}}{{c + \omega r}} 
\nonumber \\
&=& \frac{{2\pi r}}{{\left( {\frac{c}{{\left( {1 + \frac{{\omega r}}{c}} \right)}}} \right)}} - \frac{{2\pi r}}{{\left( {\frac{c}{{\left( {1 - \frac{{\omega r}}{c}} \right)}}} \right)}} = \frac{{4\pi \omega {r^2}}}{{{c^2}}}.
\label{eq:41}
\end{eqnarray}

Sag\textsubscript{2DS} is obtained by incorporating time dilation, length contraction, and differential simultaneity into Eq. (\ref{eq:39}), which are the conditions of the Franklin rT \cite{32}.  Differential simultaneity is introduced by modifying the time by adding $\mp 2 \pi \omega r\textsuperscript{2}/c\textsuperscript{2}$ to offset time with distance \cite{79}.  This reveals the isotropic one-way light speed and null Sagnac effect associated with the Franklin rT:\cite{55}
\begin{eqnarray}
{\rm{Sa}}{{\rm{g}}_{{\rm{2DS}}}} &=& \left( {\frac{{2\pi r\left( {1 - \frac{{{\omega ^2}{r^2}}}{{{c^2}}}} \right)}}{{c - \omega r}} - \frac{{2\pi \omega {r^2}}}{{{c^2}}}} \right) - \left( {\frac{{2\pi r\left( {1 - \frac{{{\omega ^2}{r^2}}}{{{c^2}}}} \right)}}{{c + \omega r}} + \frac{{2\pi \omega {r^2}}}{{{c^2}}}} \right) 
\nonumber \\
&=& \frac{{2\pi r}}{c} - \frac{{2\pi r}}{c} = 0.
\label{eq:42}
\end{eqnarray}

The relativistic Sagnac equations maintain the same form from the rotating and stationary perspectives.  Below, the same relativistic conditions as for Eqs. (\ref{eq:40})--(\ref{eq:42}) are introduced, respectively, into the Galilean stationary-frame Sagnac equation (\ref{eq:50}) to reflect effects on length and timing from the stationary perspective.  The resulting stationary-frame Sagnac equations (\ref{eq:51})--(\ref{eq:53}) are equivalent to the rotating-frame Sagnac equations (\ref{eq:40})--(\ref{eq:42}):
\begin{eqnarray}
{\rm{Sa}}{{\rm{g}}_{{\rm{1AS}}}} &=& \frac{{2\pi r\left( {\frac{c}{{c - \omega r}}} \right)\sqrt {1 - \frac{{{\omega ^2}{r^2}}}{{{c^2}}}} }}{c} - \frac{{2\pi r\left( {\frac{c}{{c + \omega r}}} \right)\sqrt {1 - \frac{{{\omega ^2}{r^2}}}{{{c^2}}}} }}{c} \nonumber \\
&=& \frac{{4\pi \omega {r^2}}}{{{c^2}\sqrt {1 - \frac{{{\omega ^2}{r^2}}}{{{c^2}}}} }},
\label{eq:51}
\end{eqnarray}
\begin{eqnarray}
{\rm{Sa}}{{\rm{g}}_{{\rm{1AS}}}} &=& \frac{{\frac{{2\pi r}}{{\sqrt {1 - \frac{{{\omega ^2}{r^2}}}{{{c^2}}}} }}\left( {\frac{c}{{c - \omega r}}} \right)\left( {1 - \frac{{{\omega ^2}{r^2}}}{{{c^2}}}} \right)}}{c} - \frac{{\frac{{2\pi r}}{{\sqrt {1 - \frac{{{\omega ^2}{r^2}}}{{{c^2}}}} }}\left( {\frac{c}{{c + \omega r}}} \right)\left( {1 - \frac{{{\omega ^2}{r^2}}}{{{c^2}}}} \right)}}{c} 
\nonumber \\
&=& \frac{{4\pi \omega {r^2}}}{{{c^2}\sqrt {1 - \frac{{{\omega ^2}{r^2}}}{{{c^2}}}} }},
\label{eq:54}
\end{eqnarray}
\begin{equation}
{\rm{Sa}}{{\rm{g}}_{{\rm{2AS}}}} = \frac{{2\pi r\left( {\frac{c}{{c - \omega r}}} \right)\left( {1 - \frac{{{\omega ^2}{r^2}}}{{{c^2}}}} \right)}}{c} - \frac{{2\pi r\left( {\frac{c}{{c + \omega r}}} \right)\left( {1 - \frac{{{\omega ^2}{r^2}}}{{{c^2}}}} \right)}}{c} = \frac{{4\pi \omega {r^2}}}{{{c^2}}},
\label{eq:52}
\end{equation}
\begin{eqnarray}
{\rm{Sa}}{{\rm{g}}_{{\rm{2DS}}}} = \left( {\frac{{2\pi r\left( {\frac{c}{{c - \omega r}}} \right)\left( {1 - \frac{{{\omega ^2}{r^2}}}{{{c^2}}}} \right)}}{c} - \frac{{2\pi \omega {r^2}}}{{{c^2}}}} \right) \nonumber \\
- \left( {\frac{{2\pi r\left( {\frac{c}{{c + \omega r}}} \right)\left( {1 - \frac{{{\omega ^2}{r^2}}}{{{c^2}}}} \right)}}{c} + \frac{{2\pi \omega {r^2}}}{{{c^2}}}} \right) = 0.
\label{eq:53}
\end{eqnarray}

\subsection{Sagnac equation two-way speeds of light} \label{section:6.3}
A given combination of forward and backward speeds of light generates a specific two-way speed of light. Thus, the one-way speeds of light associated with specific Sagnac equations generate specific rotating-frame two-way speeds of light.

The rotating-frame two-way speed of light is calculated by dividing the equivalent, rotating-frame arc lengths traversed in the two directions by the time for the light to traverse the two distances:
\begin{equation}
c{'_{{\rm{two - way}}}} = \frac{{l' + l'}}{{t{'_{{\rm{co - rot}}}} + t{'_{{\rm{counter - rot}}}}}} = \frac{{l' + l'}}{{\frac{{l'}}{{c{'_{{\rm{co - rot}}}}}} + \frac{{l'}}{{c{'_{{\rm{counter - rot}}}}}}}}.
\label{eq:44}
\end{equation}

The Sag\textsubscript{0AS} one-way speeds of light, $c \mp \omega r$, generate an anisotropic rotating-frame two-way speed of light:
\begin{equation}
c{'_{{\rm{two - waySag0AS}}}} = \frac{{l' + l'}}{{\frac{{l'}}{{c - \omega r}} + \frac{{l'}}{{c + \omega r}}}} = c\left( {1 - \frac{{{\omega ^2}{r^2}}}{{{c^2}}}} \right).
\label{eq:45}
\end{equation}

The Sag\textsubscript{1AS} one-way speeds of light, ($c \mp \omega r)/(1 - \omega \textsuperscript{2}r\textsuperscript{2}/c\textsuperscript{2})\textsuperscript{0.5}$, generate a different anisotropic two-way speed of light:
\begin{equation}
c{'_{{\rm{two - waySag1AS}}}} = \frac{{l' + l'}}{{\frac{{l'}}{{\left( {\frac{{c - \omega r}}{{\sqrt {1 - \frac{{{\omega ^2}{r^2}}}{{{c^2}}}} }}} \right)}} + \frac{{l'}}{{\left( {\frac{{c + \omega r}}{{\sqrt {1 - \frac{{{\omega ^2}{r^2}}}{{{c^2}}}} }}} \right)}}}} = c\sqrt {1 - \frac{{{\omega ^2}{r^2}}}{{{c^2}}}} .
\label{eq:46}
\end{equation}

The Sag\textsubscript{2AS} one-way speeds of light, $c/(1 \pm \omega r/c)$, generate isotropic two-way speed of light:
\begin{equation}
c{'_{{\rm{two - waySag2AS}}}} = \frac{{l' + l'}}{{\frac{{l'}}{{\left( {\frac{c}{{1 + \frac{{\omega r}}{c}}}} \right)}} + \frac{{l'}}{{\left( {\frac{c}{{1 - \frac{{\omega r}}{c}}}} \right)}}}} = c.
\label{eq:47}
\end{equation}

 The Sag\textsubscript{2DS} isotropic one-way speed of light generates isotropic two-way speed of light:
\begin{equation}
c{'_{{\rm{two - waySag2DS}}}} = \frac{{l' + l'}}{{\frac{{l'}}{c} + \frac{{l'}}{c}}} = c.
\label{eq:48}
\end{equation}

	These rotating-frame two-way speeds of light determine whether a null result is obtained in experiments that require isotropic two-way light speed to generate a null result, such as optical resonator experiments.  

\section{The Compatibility of Sagnac Equations with Optical Data} \label{section:7}
In order to determine which Sagnac equation (and its associated two-way speed of light) is compatible with optical data, we will first consider Sagnac effect data.  Data for the Sagnac effect is incompatible with the null, velocity-invariant Sag\textsubscript{2DS} equation (\ref{eq:42}).  To attempt to distinguish between the other three Sagnac equations, we will consider the highest-resolution Sagnac effect data for light, which is generated with ring laser gyroscopes that measure the Earth's rotation \cite{60}.  Ring lasers utilize differences in the beat frequency of counter-propagating light beams.  The Sagnac effect equation for beat frequency is equivalent to the conventional Sag\textsubscript{2AS} equation expressed with units that reflect beat frequency \cite{80, 81}.  The second-order differences for Sag\textsubscript{0AS}, Sag\textsubscript{1AS}, and Sag\textsubscript{2AS} are unitless quantities, and remain intact whether the Sagnac equations are expressed in units of time or beat frequency.  

Technical issues limit the resolution of ring lasers to $\Delta \omega r/\omega r\textsubscript{Earth}$ = $\sim8$ x 10\textsuperscript{-9} relative to the conventional Sag\textsubscript{2AS} beat frequency equation \cite{60}.  We will determine if this data can distinguish Sag\textsubscript{0AS} from Sag\textsubscript{2AS}, as these are the two most dissimilar equations.  Sag\textsubscript{2AS} is subtracted from Sag\textsubscript{0AS} and then the difference is divided by Sag\textsubscript{2AS} to give the ratio $\Delta$Sag\textsubscript{0AS-2AS}/Sag\textsubscript{2AS} = $1/(1- \omega\textsuperscript{2} r\textsuperscript{2}/c\textsuperscript{2}) - 1$.  This ratio is the same for Sagnac equations describing time or beat frequency.  For the peripheral velocity of Berlin (282 m/s), $\Delta$Sag\textsubscript{0AS-2AS}/Sag\textsubscript{2AS} = 8.8 x 10\textsuperscript{-13}. Because this is below the $\sim8$ x 10\textsuperscript{-9} resolution of the ring laser experiments, the data cannot distinguish between Sag\textsubscript{0AS}, Sag\textsubscript{1AS}, and Sag\textsubscript{2AS}.

Optical resonator experiments have sufficient resolution to distinguish the two-way speeds of light associated with the Sagnac equations.  Optical resonator data represents the two-way speed of light in rotating frames because the experiments are carried out on the rotating surface of the Earth, and the light path is aligned with the Earth's rotation at two points during the rotation of the apparatus.  There are eight optical resonator experiments with sufficient resolutions of 1 x 10\textsuperscript{-15} or less \cite{82, 83, 84, 85, 86, 87, 88, 89}, including an optical resonator experiment in Berlin that reported $\Delta c'\textsubscript{two-way}/c$ = 9.2 $\pm$ 10.7 x 10\textsuperscript{-19} ($\sim$1 x 10\textsuperscript{-18}) \cite{89}. 
  
	The $\Delta c'\textsubscript{two-way}/c$ value for each Sagnac equation will be determined for the Earth's peripheral velocity for the experiment in Berlin. The $\Delta c'\textsubscript{two-way}/c$ for Sag\textsubscript{0AS} is obtained by taking the difference between $c$ and the $c'\textsubscript{two-way}$ for Sag\textsubscript{0AS} (\ref{eq:45}), and then dividing by $c$, to give: $\Delta c'\textsubscript{two-way}\textsubscript{Sag0AS}/c = 1 - (1 - \omega \textsuperscript{2}r\textsuperscript{2}/c\textsuperscript{2})$. For a peripheral velocity of 282 m/s, this value is 8.8 x 10\textsuperscript{-13}. A similar calculation for the $c'\textsubscript{two-way}$ value for Sag\textsubscript{1AS} (\ref{eq:46}) gives: $\Delta c'\textsubscript{two-way}\textsubscript{Sag1AS}/c = 1 - (1 - \omega \textsuperscript{2}r\textsuperscript{2}/c\textsuperscript{2})\textsuperscript{0.5}$. For a peripheral velocity of 282 m/s, this value is 4.4 x 10\textsuperscript{-13}. Both of these divergences from $c$ would have been readily detected by the resolutions of the optical resonator experiments \cite{82, 83, 84, 85, 86, 87, 88, 89}. In contrast, the Sag\textsubscript{2AS} and Sag\textsubscript{2DS} $c'\textsubscript{two-way}$ value of $c$  (\ref{eq:47}) and (\ref{eq:48}) is compatible with the optical resonator data to the resolutions of the experiments.  

Thus, Sag\textsubscript{2AS} is compatible with both Sagnac effect and optical resonator experimental data, while Sag\textsubscript{0AS}, Sag\textsubscript{1AS}, and Sag\textsubscript{2DS} are each incompatible with one of these types of data.

\section{The Value of \boldsymbol{$\epsilon(v)$} in Rotating Frames} \label{section:8}
\subsection{\boldsymbol{$\epsilon (v)$} is a function of the slope of the \boldsymbol{$x'$}-axis} \label{C}
We have previously shown that the value of $\epsilon (v)$ is a function of the slope of the $x'$-axis \cite{8}.  Here we briefly describe this derivation. In spacetime diagrams, the $x'$-axis is described in ``stationary'' coordinates by the line for which the $t'$ values are 0 (see Fig. 1). Therefore, to describe $\epsilon (v)$ in terms of the slope of the $x'$-axis ($m$): $t'$ is set to 0 in the MS time equation (\ref{eq:20}), which makes the $x$ and $t$ values refer to the $x'$-axis; and $x'$ is substituted with its value from Eq. (\ref{eq:21}), to give:
\begin{equation}
0 = at + \epsilon b(x - vt).
\label{eq:55}
\end{equation}
Substituting $m$ for $t/x$, and the empirical relativistic values for $a(v)$ and $b(v)$ gives:
\begin{equation}
\epsilon  = \frac{{ - m\left( {1 - \frac{{{v^2}}}{{{c^2}}}} \right)}}{{1 - mv}}.
\label{eq:56}
\end{equation}
Thus, the value of $\epsilon (v)$ is determined by the slope of the $x'$-axis, which determines the simultaneity framework.  Rearranging Eq. (\ref{eq:56}) shows the value of $m$ in terms of $\epsilon (v)$:
\begin{equation}
m = \frac{\epsilon }{{\epsilon v - \left( {1 - \frac{{{v^2}}}{{{c^2}}}} \right)}}.
\label{eq:57}
\end{equation}

\subsection{The $\epsilon (v)$ value from unidirectional one-way light signals} \label{section:8.1}
We have previously shown that the value of $\epsilon (v)$ can be determined from the anisotropy of light signals sent in two directions \cite{8}.  Here we describe how the value of $\epsilon (v)$ can be determined from the anisotropy of a unidirectional light signal.  We define an anisotropy signal ($AS$) as the difference between the observed timing of a one-way light signal in the ``moving'' frame and the timing expected for isotropic light speed. $AS$ is derived using a geometric proof in \ref{E}.  The equation for $AS$ is:
\begin{equation}
AS = \frac{{l\left( {\frac{v}{{{c^2}}} - m} \right)}}{{1 - mv}},
\label{eq:70}
\end{equation}
where $m$ is the slope of the $x'$-axis.  Rearranging Eq. (\ref{eq:70}) gives:
\begin{equation}
m = \frac{{\frac{{lv}}{{{c^2}}} - AS}}{{l - vAS}}.
\label{eq:71}
\end{equation}

As derived in \ref{E}, $l$ represents the distance ``at rest'' in the ``stationary'' frame that is the counterpart of the distance that the light spans in the ``moving'' frame.  For example, when light spans a ``moving'' platform of length of $l'=1$, the corresponding value for the platform ``at rest'' in the ``stationary'' frame is $l = 1$.  Because the two length terms have equivalent numbers of distance units, Eqs. (\ref{eq:70}) and (\ref{eq:71}) could be expressed with $l'$ rather than $l$, and the same values would be obtained.  Because $l$ in Eqs. (\ref{eq:70}) and (\ref{eq:71}) refers to the non-``moving''/non-rotating length, a light path of one circumference is unambiguously $2 \pi r$.

To describe $\epsilon (v)$ in terms of $AS$, the values of $m$ from Eqs. (\ref{eq:71}) and (\ref{eq:57}) are set to equal each other and then simplified: 
\begin{equation}
\epsilon  = \frac{{AS}}{l} - \frac{v}{{{c^2}}}.
\label{eq:72}
\end{equation}
Thus, the value of $\epsilon (v)$ can be determined from the one-way light speed anisotropy. 

\subsection{Calculating the value of $\epsilon(v)$ in rotating frames} \label{section:8.2}
The unidirectional Sagnac effect (Sagnac\textsubscript{UD}) describes the extent of light speed anisotropy in one direction around the circumference of a rotating disk.  The Sagnac\textsubscript{UD} of Sag\textsubscript{2AS} (in the co-rotating direction) is:
\begin{equation}
{\rm{Sagna}}{{\rm{c}}_{{\rm{UD}}}} = \frac{{2\pi \omega {r^2}}}{{{c^2}}}.
\label{eq:76}
\end{equation}
This equation is equivalent to the ``Sagnac correction'', which describes the light speed anisotropy that is observed in GPS communications \cite{90}.

The value of $\epsilon (v)$ in terms of $AS$ (Eq. \ref{eq:72}) expressed with polar coordinates is:
\begin{equation}
\epsilon  = \frac{{AS}}{l} - \frac{{\omega r}}{{{c^2}}}.
\label{eq:73}
\end{equation}
Sagnac\textsubscript{UD} (Eq. \ref{eq:76}) is the empirical value of $AS$ that describes one-way light speed anisotropy in rotating frames.  The value of $l$ for one circumference is $2 \pi r$ (see Sec. \ref{section:8.1}).  Substitution of these values into Eq. (\ref{eq:73}) gives a null $\epsilon (v)$ value:
\begin{equation}
\epsilon  = \frac{{2\pi \omega {r^2}}}{{2\pi r{c^2}}} - \frac{{\omega r}}{{{c^2}}} = 0.
\label{eq:74}
\end{equation}
A similar analysis with the revised Robertson test theory shows that its simultaneity parameter is also null in rotating frames (see \ref{H}).

Experimental observations support a null value of $\epsilon (v)$ within rotating frames.  A null $\epsilon (v)$ predicts one-way light speed anisotropy, the Sagnac effect, and directional time dilation (see Sec. \ref{section:3.2} and Fig. 1(b)).  One-way light speed anisotropy and the Sagnac effect are observed in rotating frames \cite{56, 58, 59}.  Furthermore, time dilation occurs in rotating frames in an absolute and directional manner, as shown by direct comparisons of atomic clocks before and after flights \cite{34, 35, 91}.

\section{The Empirical MS Parameter Values Generate the ALT rT} \label{section:9}
When the values of all the parameters for a kinematic test theory are empirically determined, those values can be inserted into the test theory equations to generate the transformation that is compatible with the empirical data. With the results in Sec. \ref{section:8.2}, the empirical values for all of the MS parameters are now known for rotating frames.  Incorporating these values into the MS test theory equations (\ref{eq:20})--(\ref{eq:23}) generates the following transformation equations:
\begin{equation}
dt' = dt\sqrt {1 - \frac{{{\omega ^2}{r^2}}}{{{c^2}}}},
\label{eq:16}
\end{equation}
\begin{equation}
d\theta 'r' = \frac{{d\theta r - \omega rdt}}{{\sqrt {1 - \frac{{{\omega ^2}{r^2}}}{{{c^2}}}} }},
\label{eq:17}
\end{equation}
\begin{equation}
dr' = dr,
\label{eq:18}
\end{equation}
\begin{equation}
dz' = dz.
\label{eq:19}
\end{equation}
These equations define the ALT rT \cite{33}.  This implies that the ALT rT accurately reflects relativistic effects in rotating frames. 

\section{Conclusions}\label{section:11}
Our study focuses on the simultaneity framework in rotating frames.  This area of research has historically not been settled, with publications that support either differential simultaneity or absolute simultaneity in rotating frames.  A number of studies suggest that locally co-moving inertial frames, each of which manifest differential simultaneity, will generate an overt Sagnac effect when integrated over a full rotation, suggesting that differential simultaneity is present in rotating frames \cite{64, 65, 66, 67, 92}.  Other studies support differential simultaneity in rotating frames by suggesting that the time gap (which results from differential simultaneity) generates the observed Sagnac effect \cite{61, 93, 94, 95, 96, 97}.  The time gap arises when time is offset with distance over a circumference, which generates a time discontinuity between the adjacent starting and ending points that were used for calculating the time offset. Our companion study assesses these two mechanisms using high-resolution optical data, and comes to the conclusion that neither mechanism matches the data \cite{55}.  Countervailing studies have suggested that rotating frames exhibit absolute simultaneity.  These studies have indicated that the differential-simultaneity-associated time gap introduces invalidating inconsistencies when applied to rotating frames or cylindrical spacetime, while absolute simultaneity is fully consistent with rotating-frame data \cite{23, 25, 33, 42, 43, 44}.  Our study supports the conclusion that rotating frames exhibit absolute simultaneity.

Our study uses the MS test theory to directly assess the simultaneity framework in rotating frames.  The MS test theory is the most widely used kinematic relativistic test theory.  The empirical values of three MS parameters have been known for decades \cite{5, 6, 7}.  However the $\epsilon (v)$ parameter was considered a convention on clock synchronization and was not assigned an empirical value.  In Sec. \ref{section:3}, we show that $\epsilon (v)$ is not a convention.  In Sec. \ref{C}, we show that $\epsilon (v)$ is a descriptor of the simultaneity framework whose value reflects the extent to which time is offset with distance \cite{8}.  In Sec. \ref{section:8.1}, we show that the value of $\epsilon (v)$ can be determined from the extent of anisotropy of the unidirectional one-way speed of light, which in rotating frames is described by the Sagnac effect.  As described in Sec. \ref{section:1.3}, the use of the Sagnac effect to calculate $\epsilon (v)$ removes the confounding issue of the method of clock synchronization because the Sagnac effect does not involve clock synchronization.  To determine which Sagnac equation to utilize, we analyzed the published Sagnac equations.  In Sec. \ref{section:6}, we show that the published Sagnac equations form a relativistic series, and that Sagnac equations are equivalent from the rotating and stationary perspectives.  In Sec. \ref{section:7}, we show that only the conventional Sagnac equation, Sag\textsubscript{2AS}, is compatible with both Sagnac effect and optical resonator data.  In Sec. \ref{section:8}, we use Sag\textsubscript{2AS} to determine that $\epsilon (v)$ has a null value in rotating frames. 

Incorporation of the empirical values of the MS test theory parameters into the test theory equations generates the ALT rT (Sec. \ref{section:9}).  This implies that the ALT rT accurately describes relativistic kinematics in rotating frames.  The ALT rT is the only major kinematic rT that is compatible with a broad range of rotational relativistic data, including predicting the conventional Sagnac effect and isotropic two-way speed of light (see our companion study \cite{55}).  

The null value of $\epsilon (v)$ in rotating frames implies that time is not offset based on distance.  The null value for $\epsilon (v)$ was obtained using the Sagnac effect description of anisotropy in the rotating-frame one-way speed of light.  Therefore, the null value of $\epsilon (v)$ in rotating frames should be valid at all distance scales for which the Sagnac effect is observed, which currently spans radii from 500 $\mu$m \cite{98} to that of the Earth \cite{68}.  Given the current spatial limitations of the validation, there are no empirical disagreements between the ALT rT and dissimilar, proposed transformations that operate on much smaller distance scales to describe relativistic effects in response to the rotations of subatomic particles \cite{99}.

\appendix

\section{Sagnac Effect Lengths from the Stationary Perspective}\label{A}
Here we derive the arc lengths for light propagation for the Sagnac effect from the stationary perspective in a Galilean framework that lacks relativistic effects and has absolute simultaneity.  

From the stationary perspective, light propagates at the isotropic one-way speed of light, $c$ \cite{5}. Figure \ref{FigC.1} provides diagrams of the light paths of light sent in the co-rotating and counter-rotating directions in a Sagnac effect experiment. In the co-rotating direction, the light signal is emitted from the emitter/receiver at position E\textsubscript{co-rot} and received by the emitter/receiver at position R. The arc length that the emitter/receiver rotates between emission and reception is denoted $x\textsubscript{co-rot}$. The arc length of light propagation is denoted $LP\textsubscript{co-rot}$ (Fig. \ref{FigC.1}), and is:

\begin{figure}[t]
\centering\includegraphics[width=4.0in]{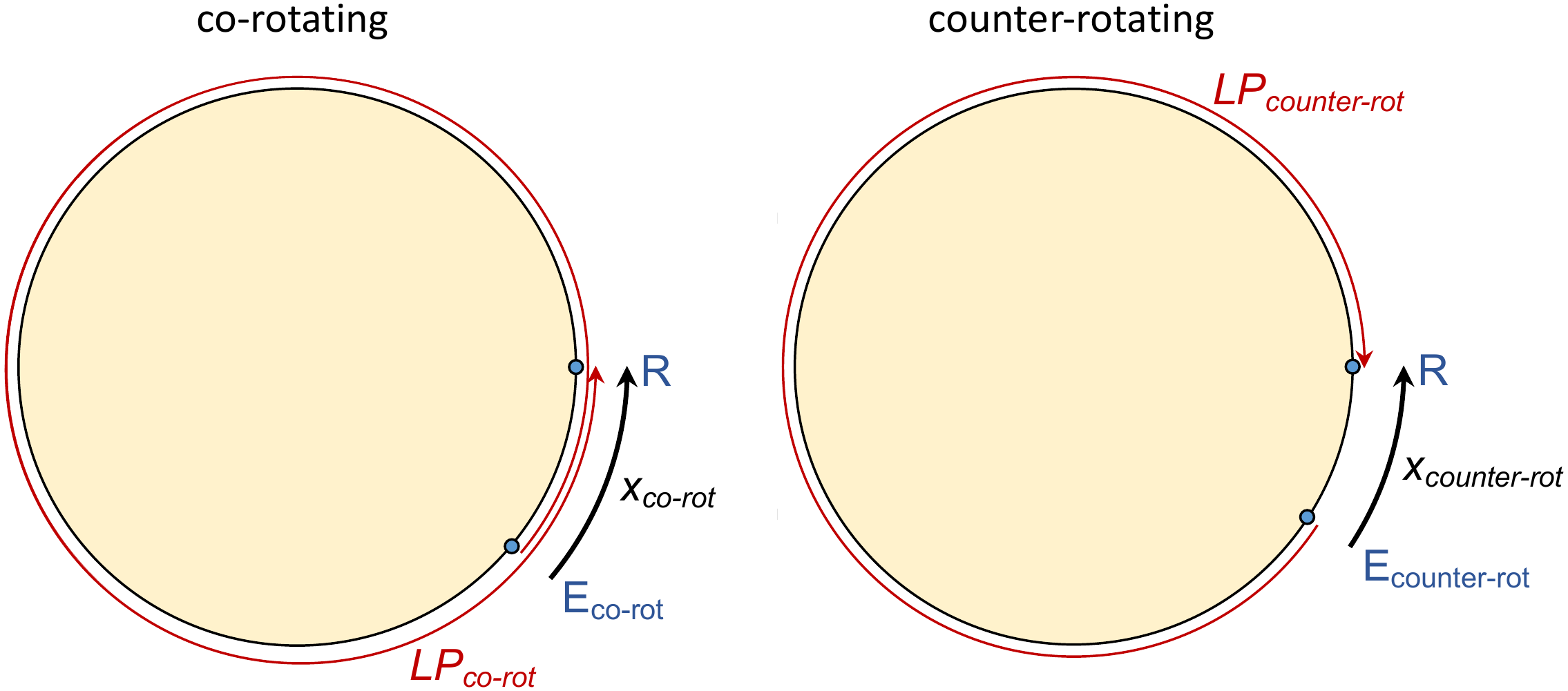}
\caption{Diagrams of the light paths in a Sagnac effect experiment from the stationary perspective for co-rotating (left) and counter-rotating (right) light signals. The rotating disk is shaded yellow. The peripheral velocity of the rotating disk is 0.1$c$. In both diagrams, the emitter/receiver is represented as a blue dot. The emitter/receiver is shown at the position of light signal emission (E) and reception (R). The thin red line with an arrow illustrates the light path from emission to reception; this corresponds to the arc length for light propagation, $LP$. The thick black line with an arrow is the arc length that the emitter/receiver rotates between light signal emission and reception ($x$). Values that differ between the two diagrams are labeled as co-rotating or counter-rotating.}
\label{FigC.1}
\end{figure}

\begin{equation}
L{P_{co - rot}} = 2\pi r + {x_{co - rot}}.
\label{C.10}
\end{equation}
The time that the light signal propagates is $LP\textsubscript{co-rot}$ divided by $c$:
\begin{equation}
{t_{co - rot}} = \frac{{2\pi r + {x_{co - rot}}}}{c}.
\label{C.1}
\end{equation}
The time that the emitter/receiver takes to rotate from point E\textsubscript{co-rot} to point R is:
\begin{equation}
{t_{co - rot}} = \frac{{{x_{co - rot}}}}{{\omega r}}.
\label{C.2}
\end{equation}
The time interval between light emission and reception are equivalent for Eqs. (\ref{C.1}) and (\ref{C.2}). Setting these two equations equal to each other and solving for $x\textsubscript{co-rot}$ gives:
\begin{equation}
{x_{co - rot}} = \frac{{2\pi r}}{{\left( {\frac{c}{{\omega r}} - 1} \right)}}.
\label{C.3}
\end{equation}
Incorporating the value of $x\textsubscript{co-rot}$ (\ref{C.3}) into Eq. (\ref{C.10}) gives:
\begin{equation}
L{P_{co - rot}} = 2\pi r\left( {1 + \frac{1}{{\left( {\frac{c}{{\omega r}} - 1} \right)}}} \right).
\label{C.4}
\end{equation}
The terms in parentheses in Eq. (\ref{C.4}) can be rearranged to give the ratio $c/c'\textsubscript{co-rot}$, thus giving the value of $LP\textsubscript{co-rot}$:
\begin{equation}
L{P_{co - rot}} = 2\pi r\left( {\frac{c}{{c - \omega r}}} \right).
\label{C.5}
\end{equation}

In the counter-rotating direction, $LP\textsubscript{counter-rot}$ (Fig. \ref{FigC.1}) is:
\begin{equation}
L{P_{counter - rot}} = 2\pi r - {x_{counter - rot}}.
\label{C.11}
\end{equation}
The time that the light signal propagates is $LP\textsubscript{counter-rot}$ divided by $c$:
\begin{equation}
{t_{counter - rot}} = \frac{{2\pi r - {x_{counter - rot}}}}{c}.
\label{C.6}
\end{equation}
The time that the emitter/receiver takes to rotate from E\textsubscript{counter-rot} to R is:
\begin{equation}
{t_{counter - rot}} = \frac{{{x_{counter - rot}}}}{{\omega r}}.
\label{C.7}
\end{equation}
Setting Eqs. (\ref{C.6}) and (\ref{C.7}) equal to each other and solving for $x\textsubscript{counter-rot}$ gives:
\begin{equation}
{x_{counter - rot}} = \frac{{2\pi r}}{{\left( {\frac{c}{{\omega r}} + 1} \right)}}.
\label{C.8}
\end{equation}
Incorporating the value of $x\textsubscript{counter-rot}$ (\ref{C.8}) into Eq. (\ref{C.11}) gives:
\begin{equation}
L{P_{counter - rot}} = 2\pi r\left( {1 - \frac{1}{{\left( {\frac{c}{{\omega r}} + 1} \right)}}} \right).
\label{C.9}
\end{equation}
The terms in parentheses in Eq. (\ref{C.9}) can be rearranged to give the ratio $c/c'\textsubscript{counter-rot}$, thus giving the value of $LP\textsubscript{counter-rot}$:
\begin{equation}
L{P_{counter - rot}} = 2\pi r\left( {\frac{c}{{c + \omega r}}} \right).
\label{C.5}
\end{equation}

\section{Derivation of the Anisotropy Signal \boldsymbol{$AS$}}\label{E}
\begin{figure}[t]
\centering\includegraphics[width=3.0in]{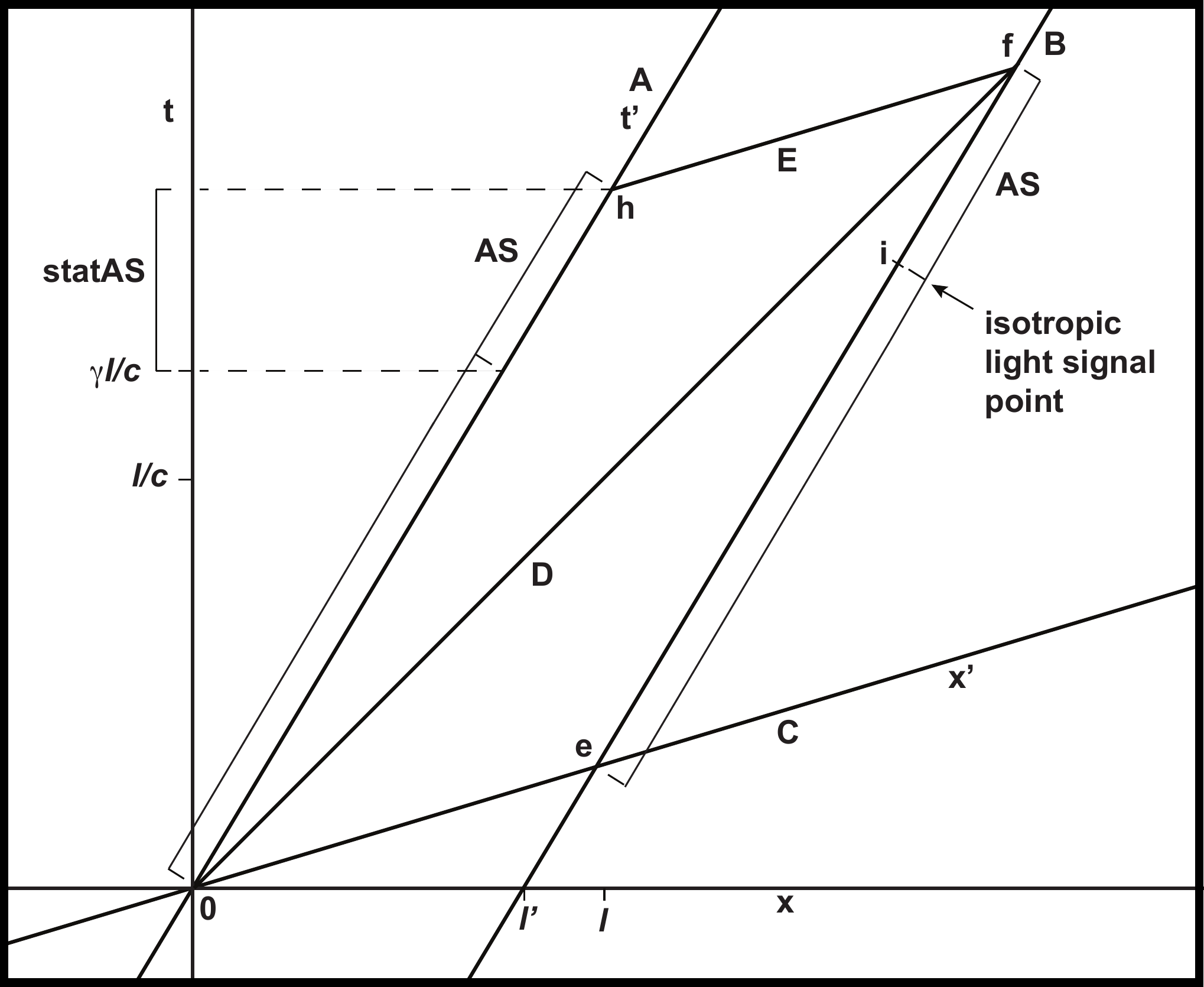}
\caption{Spacetime diagram with $v = 0.6c$.  Lines are labeled in capital letters.  Points are labeled in lowercase letters. Line D represents a light signal sent a distance of $l'$ from the origin (reaching point $f$ on line B).  $AS$ represents the difference between the timing of an observed light signal and an isotropic light signal.  On line B: the ``moving'' time for the total light signal to go from the origin to point $f$ is from point $e$ to point $f$; the time for an isotropic light signal is from $e$ to $i$; and the time for $AS$ is from $i$ to $f$.  The equivalent timing is shown on line A.  $statAS$ is the time for $AS$ in the ``stationary'' frame and is shown by a bracket with dashed lines.}
\label{FigE.1}
\end{figure}

Here, we geometrically derive the equation for the anisotropy signal $AS$. $AS$ is the difference between the observed time that light takes to propagate a length $l'$ and the time that it would take if light speed were isotropic.  The spacetime diagram in Fig. \ref{FigE.1} has normal time dilation and length contraction, and depicts a slope of the $x'$-axis that is intermediate between that of the LT and ALT.  A light signal is sent from the origin to a distance of $l'$, where it is received at point $f$ on line B.  

	In the geometric derivation, $l$ refers to the non-moving ``stationary''-frame distance that corresponds to the ``moving''-frame $l'$ distance.  As an example, if light is sent between two ends of a ``moving'' platform of length $l' = 1$, this corresponds to $l = 1$, for the length of the platform when it is not moving.

Line A is the $t'$-axis, the equation is:
\begin{equation}
t = \frac{x}{v}.
\label{A.1}
\end{equation}
Line B is parallel to the $t'$-axis and intersects the $x$-axis at length $l'$:
\begin{equation}
t = \frac{x}{v} - \frac{l}{{\gamma v}},
\label{A.2}
\end{equation}
where 
\begin{equation}
\gamma  = \frac{1}{{\sqrt {1 - \frac{{{v^2}}}{{{c^2}}}} }}.
\label{A.3}
\end{equation}
Line C is the $x'$-axis:
\begin{equation}
t = mx.
\label{A.4}
\end{equation}
Line D corresponds to the path of a light signal sent from the origin:
\begin{equation}
t = \frac{x}{c}.
\label{A.5}
\end{equation}
The $x$ and $t$ coordinates of point $f$ (at the intersection of lines E, D, and B) are:
\begin{equation}
{x_f} = \frac{{lc}}{{\gamma (c - v)}},{t_f} = \frac{l}{{\gamma (c - v)}}.
\label{A.6}
\end{equation}
Line E is parallel to the $x'$-axis, and intersects points $h$ and $f$:
\begin{equation}
t = mx + \frac{{l(1 - mc)}}{{\gamma (c - v)}}.
\label{A.7}
\end{equation}
Point $h$ is the intersection of lines A and E. The $t$ coordinate of point $h$ is:
\begin{equation}
{t_h} = \frac{{l(1 - mc)}}{{\gamma (c - v)(1 - mv)}}.
\label{eq:A.8}
\end{equation}

	The segment of line B from point $e$ to point $i$ corresponds to the ``moving''-frame time that a light signal sent from the origin at time 0 would take to reach a length of $l'$ if light speed were isotropic.  In ``stationary'' time units, the length of the segment of line B from point $e$ to point $i$ is $\gamma l/c$.  The isotropic time in ``stationary'' units is $\gamma l/c$ because time dilation causes the ``moving'' time units to extend further relative to ``stationary'' units on a spacetime diagram, so that a ``moving'' time of $l'/c$ extends over a ``stationary'' time of $\gamma l/c$ (see Fig. 1(a); and for a mathematical treatment \cite{47}).  

	The ``moving''-frame time that it takes the light to reach a length $l'$ is from point $e$ to point $f$ ($e$ is at time $t' = 0$, and $f$ is when the light strikes line B).  $AS$ is the difference in ``moving''-frame times between how long it takes light to reach a length of $l'$ (from point $e$ to $f$ on line B) and the time it would take if light were isotropic (from point $e$ to $i$ on line B).  Thus, $AS$ is the time from point $i$ to point $f$ on line B.

	The segment of line A from the origin to the ``stationary'' time $\gamma l/c$ is equivalent to the distance from point $e$ to point $i$ on line B.  The segment of line A from that point to point $h$ is equivalent to $AS$ in the ``moving'' frame, and to ``stationary'' $AS$ ($statAS$) in the ``stationary'' frame.  $statAS$ is therefore equivalent to the time at point $h$, $t\textsubscript{h}$ (\ref{eq:A.8}), minus the time $\gamma l/c$:
\begin{equation}
statAS = \frac{{l(1 - mc)}}{{\gamma (c - v)(1 - mv)}} - \frac{{\gamma l}}{c}.
\label{eq:A.9}
\end{equation}
Simplifying Eq. (\ref{eq:A.9}) gives:
\begin{equation}
statAS = \frac{{\gamma l\left( {\frac{v}{{{c^2}}} - m} \right)}}{{1 - mv}}.
\label{eq:A.10}
\end{equation}
$AS$ is obtained by multiplying $statAS$ by $1/\gamma$, reflecting the reduced period of time in the ``moving'' frame relative to the ``stationary'' frame.  $AS$ is therefore:
\begin{equation}
AS = \frac{{l\left( {\frac{v}{{{c^2}}} - m} \right)}}{{1 - mv}}.
\label{eq:A.11}
\end{equation}

\section{The Revised Robertson Test Theory Simultaneity Parameter}\label{H}
The Robertson test theory \cite{9} is constrained in its analysis of the simultaneity framework.  The Robertson test theory explicitly incorporates Einstein synchronization \cite{4, 9}, which restricts the framework to differential simultaneity.  The Robertson test theory has the following equations:
\begin{equation}
t = {a_0}t' + {a_1}\left( {\frac{v}{{{c^2}}}} \right)x',
\label{eq:B.1}
\end{equation}
\begin{equation}
x = {a_0}vt' + {a_1}x',
\label{eq:B.2}
\end{equation}
\begin{equation}
y = {a_2}y',
\label{eq:B.3}
\end{equation}
\begin{equation}
z = {a_2}z'.
\label{eq:B.4}
\end{equation}

The test theory is unable to describe absolute simultaneity because that would require a null value for $a\textsubscript{1}(v)$ in (\ref{eq:B.1}) to eliminate the $x'$ term that offsets time with distance.  However, a null value of $a\textsubscript{1}(v)$ is not possible because of its use in (\ref{eq:B.2}). 

The revised Robertson test theory eliminates the constraints on simultaneity of the original theory \cite{100}. The revised Robertson test theory equations are:
\begin{equation}
t' = jt + hx,
\label{eq:B.9}
\end{equation}
\begin{equation}
x' = a\left( {x - vt} \right),
\label{eq:B.10}
\end{equation}
\begin{equation}
y' = ey,
\label{eq:B.11}
\end{equation}
\begin{equation}
z' = ez.
\label{eq:B.12}
\end{equation}
$h(v)$ is the simultaneity parameter because it modifies the distance component in the time equation (\ref{eq:B.9}), thus describing how time is offset with distance. 

	To determine the value of $h(v)$ in rotating frames, we will derive the equation linking $h(v)$ to the anisotropy signal $AS$. We showed in Eq. (\ref{eq:71}) that the slope of the $x'$-axis ($m$) is a function of $AS$:
\begin{equation}
m = \frac{{\frac{{lv}}{{{c^2}}} - AS}}{{l - vAS}}.
\label{eq:B.17}
\end{equation}
Parameter $j(v)$ with experimentally-determined relativistic terms \cite{100} is:
\begin{equation}
j = \sqrt {1 - \frac{{{v^2}}}{{{c^2}}}}  - hv.
\label{eq:B.18}
\end{equation}
The value of $m$ in revised Robertson parameters can be obtained by setting $t' = 0$ in Eq. (\ref{eq:B.9}) (to represent the $x'$-axis) and solving for $m$ ($t/x$):
\begin{equation}
m =  - \frac{h}{j}.
\label{eq:B.19}
\end{equation}
Substituting the values of $m$ from Eq. (\ref{eq:B.17}), and $j(v)$ from Eq. (\ref{eq:B.18}), into Eq. (\ref{eq:B.19}), and then solving for $h(v)$ gives:
\begin{equation}
h = \frac{{\frac{{AS}}{l} - \frac{v}{{{c^2}}}}}{{\sqrt {1 - \frac{{{v^2}}}{{{c^2}}}} }}.
\label{eq:B.20}
\end{equation}

Substituting the Sagnac\textsubscript{UD} of Sag\textsubscript{2AS} (\ref{eq:76}), which describes one-way light speed anisotropy, for $AS$ and expressing the formula with polar coordinates gives a null value for $h(v)$ in rotating frames:
\begin{equation}
h = \frac{{\frac{{2\pi \omega {r^2}}}{{2\pi r{c^2}}} - \frac{{\omega r}}{{{c^2}}}}}{{\sqrt {1 - \frac{{{\omega ^2}{r^2}}}{{{c^2}}}} }} = 0.
\label{eq:B.21}
\end{equation}

The empirical values for the revised Robertson parameters are: $a(v) = 1/(1-v\textsuperscript{2}/c\textsuperscript{2})\textsuperscript{0.5}$; $j(v)=(1-v\textsuperscript{2}/c\textsuperscript{2})\textsuperscript{0.5}$; and $e(v)=1$ \cite{100}; and $h(v)=0$.  Substitution of these values into Eqs. (\ref{eq:B.9})--(\ref{eq:B.12}) generates the ALT rT equations (\ref{eq:16})--(\ref{eq:19}), as the corresponding transformation.

\section*{Acknowledgments}

We thank Y. Kipreos for helpful discussions.

\end{document}